%% file: ms.tex
\documentclass[10pt,preprint2]{aastex}
\shorttitle{Search for planets in NGC~6791}
\shortauthors{Mochejska et al.}
\begin{document}

\title{Planets in Stellar Clusters Extensive Search. III.~A search for
transiting planets in the metal-rich open cluster
NGC~6791.\altaffilmark{1}}
\altaffiltext{1}{Based on data from the FLWO 1.2m telescope}

\author{B.~J.~Mochejska\altaffilmark{2}}

\affil{Purdue University, Department of Physics, 525 Northwestern
Ave., West Lafayette, IN~47907}
\email{bmochejs@cfa.harvard.edu}
\author{K.~Z.~Stanek, D.~D.~Sasselov, A.~H.~Szentgyorgyi,
G.~{\'A}.~Bakos\altaffilmark{2,3}, J.~Devor, V.~Hradecky, D.~P.~Marrone,
J.~N.~Winn\altaffilmark{2} \& M.~Zaldarriaga}
\affil{Harvard-Smithsonian Center for Astrophysics, 60 Garden St.,
Cambridge, MA~02138}
\email{kstanek, sasselov, saint, gbakos, jdevor, vhradecky,
dmarrone, jwinn, mzaldarriaga@cfa.harvard.edu}
\altaffiltext{2}{Hubble Fellow}
\altaffiltext{3}{Also at Konkoly Observatory}

\begin{abstract}
We have undertaken a long-term project, Planets in Stellar Clusters
Extensive Search (PISCES), to search for transiting planets in open
clusters. In this paper we present the results for NGC~6791 -- a very
old, populous, metal rich cluster. We have monitored the cluster for
over 300 hours, spread over 84 nights. We have not detected any good
transiting planet candidates. Given the photometric precision and
temporal coverage of our observations, and current best estimates for
the frequency and radii of short-period planets, the expected number
of detectable transiting planets in our sample is 1.5. We have
discovered 14 new variable stars in the cluster, most of which are
eclipsing binaries, and present high precision light curves, spanning
two years, for these new variables and also the previously known
variables.
\end{abstract}

\keywords{planetary systems -- binaries: eclipsing -- cataclysmic
variables -- stars: variables: other -- color-magnitude diagrams }

\section{{\sc Introduction}}
We have undertaken a long-term project, Planets in Stellar Clusters
Extensive Search (PISCES), to search for transiting planets in open
clusters. To date we have published a feasibility study based on one
season of data for NGC~6791 (Mochejska et al.\ 2002, hereafter
Paper~I). We have also presented a variable star catalog in our second
target, NGC~2158, based on the data from the first observing season
(Mochejska et al.\ 2004, hereafter Paper~II).

In this paper we present the results of a search for transiting
planets in the open cluster NGC~6791
$[(\alpha,\delta)_{2000}=(19^h20.8^m, +37^{\circ}51')]$. It is a very
populous (Kaluzny \& Udalski 1992), very old ($\tau$=8 Gyr), extremely
metal rich ([Fe/H]=+0.4) cluster, located at a distance modulus of
(m-M)$_V$ = 13.42 (Chaboyer, Green \& Liebert 1999).

Stars hosting planets are known to be, on the average, significantly
more metal rich than those without (Santos et al.\ 2001, 2004).  Two
scenarios have been proposed to explain this phenomenon.  Some studies
favor a ``primordial'' metallicity enhancement, i.e.\ reflecting the
original metallicity of the gas from which the star formed (Santos et
al.\ 2004; Pinsonneault et al.\ 2001). In this scenario planet
formation would be more prolific in a metal-rich environment (Ida \&
Lin 2004). Others suggest that the host stars were enriched by the
infall of other giant gas planets (Lin 1997) or small planetary bodies
like asteroids (Murray \& Chaboyer 2002).

The observed lack of planets in the core (Gilliland et al.\ 2000) and
the uncrowded outer regions (Weldrake et al.\ 2005) of the low
metallicity ([Fe/H]$=-0.7$) globular cluster 47 Tuc suggests that the
source of the metallicity enhancement in planet hosts is most likely
``primordial''. Open clusters offer the possibility of observing a
large number of stars with the same, known a priori
metallicity. NGC~6791, with its high metallicity and large number of
stars, seems particularly well suited as a target for transiting
planet search.

Targeting open clusters also eliminates the problem of false
detections due to blended eclipsing binary stars, which are a
significant contaminant in the Galactic field searches (over 90\% of
all candidates; Konacki et al.\ 2003; Udalski et al.\ 2002a, 2002b).
Blending causes a large decrease of the depth of the eclipses and
mimics the transit of a much smaller object, such as a planet. As
opposed to dense star fields in the disk of our Galaxy, open clusters
located away from the galactic plane are sparse enough for blending to
be negligible.

There are two key elements in a survey for transiting planets. The
most commonly emphasized requirement is the high photometric
precision, at the 1\% level. The more often overlooked factor is the
need for very extensive temporal coverage.

Extensive temporal coverage is important because even for planets with
periods between 1 and 2 days, the fractional transit length is only
$\sim$5\% of the period, and it drops to $\sim$2\% for periods 2-10
days. During the remaining 95-98\% of the period the system is
photometrically indistinguishable from stars without transiting
planets.  To our best knowledge, PISCES is the most extensive search
for transiting planets in open clusters in terms of temporal coverage
with a 1 m telescope.

NGC~6791 has been previously searched for transiting planets by Bruntt
et al.\ (2003), who found three transit-like events and seven other
lower probability events which may possibly be due to instrumental
effects. Of the three best candidates, none exhibited more than one
transit and only one is located on the cluster main sequence. Bruntt
et al.\ (2003) used the 2.5 m NOT telescope, which allowed them to
obtain higher photometric precision and denser time sampling, but
their temporal coverage was much inferior to ours: $\sim$24 hours
spread over 7 nights, compared to our $>300$ hours, collected over 84
nights.

The paper is arranged as follows: \S 2 describes the observations, \S
3 summarizes the reduction procedure, \S 4 outlines the search
strategy for transiting planets, \S 5 gives an estimate of the
expected number of transiting planet detections, \S 6 describes the
candidates previously reported by Bruntt et al.\ (2003) and \S 7
contains the variable star catalog. Concluding remarks are found in \S
8.

\section{{\sc Observations}} 
\begin{figure}[t]
\plotone{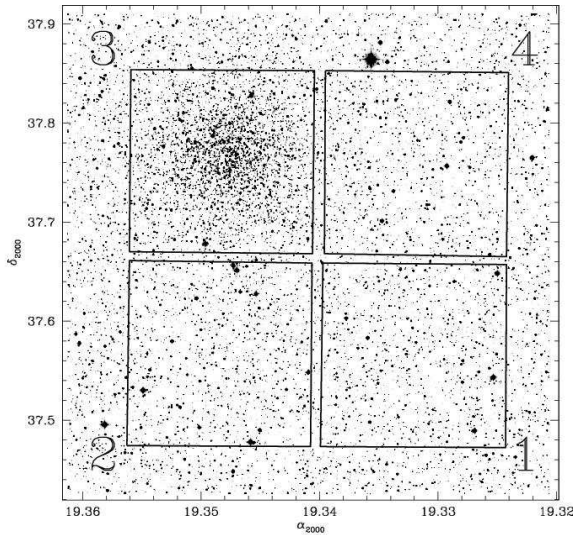}
\caption{Digital Sky Survey image of NGC~6791 showing the field of
view of the 4Shooter. The chips are numbered clockwise from 1 to 4
starting from the bottom left chip. NGC~6791 is centered on Chip~3.
North is up and east is to the left.}
\label{chips}
\end{figure}

The data analyzed in this paper were obtained at the Fred Lawrence
Whipple Observatory (FLWO) 1.2 m telescope using the 4Shooter CCD
mosaic with four thinned, back side illuminated AR coated Loral
$2048^2$ CCDs (Szentgyorgyi et al.\ in preparation).  The camera, with
a pixel scale of $0\farcs 33$ pixel$^{-1}$, gives a field of view of
$11\farcm 4\times 11\farcm 4$ for each chip. The cluster was centered
on Chip~3 (Fig.~\ref{chips}). The data were collected during 84
nights, from 2001 July 9 to 2003 July 10. A total of $1118\times
900$~s $R$ and $233\times 450$~s $V$-band exposures were obtained. The
$V$-band dataset was supplemented with $93\times 450$ s exposures
collected between 19 September 1998 and 5 November 1999 (previously
analyzed by Mochejska, Stanek \& Kaluzny 2003).

\section{{\sc Data Reduction}}

\subsection{{\it Image Subtraction Photometry}}
\label{imsub}
The preliminary processing of the CCD frames was performed with the
standard routines in the IRAF ccdproc package.\footnote{IRAF is
distributed by the National Optical Astronomy Observatories, which are   
operated by the Association of Universities for Research in Astronomy,
Inc., under cooperative agreement with the NSF.}

Photometry was extracted using the ISIS image subtraction package
(Alard \& Lupton 1998; Alard 2000), as described in detail in Paper~I.

The ISIS reduction procedure consists of the following steps: (1)  
transformation of all frames to a common $(x,y)$ coordinate grid; (2)
construction of a reference image from several of the best exposures;    
(3) subtraction of each frame from the reference image; (4) selection 
of stars to be photometered and (5) extraction of profile photometry
from the subtracted images.

All computations were performed with the frames internally subdivided
into four sections ({\tt sub\_x=sub\_y=2}). Differential brightness   
variations of the background were fit with a second degree polynomial
({\tt deg\_bg=2}). A convolution kernel varying quadratically with
position was used ({\tt deg\_spatial=2}). The psf width ({\tt
psf\_width}) was set to 33 pixels and the photometric radius ({\tt   
radphot}) to 5 pixels. The reference images were constructed from
25 best exposures in $R$ and 16 in $V$.

We slightly modified the reduction pipeline described in Paper~I by
introducing a procedure to remove photometry from epochs where a star
was located in the proximity of bad columns. This task was somewhat
complicated by the fact that the original {\tt interp} program uses
spline functions to remap each image to the template's $(x,y)$
coordinate grid. If an image is masked before transformation, masked
regions will spread over adjacent columns in the remapped image. To
avoid this problem, we performed a linear transformation of the bad
pixel masks for each image using the coefficients output by the {\tt
fitn} program.  The shifted masks were applied to subtracted
images. The {\tt Cphot} program was modified, so that it ignored
epochs where a bad pixel was within {\tt radphot} pixels of a star's
centroid.

\begin{figure}[t]
\plotone{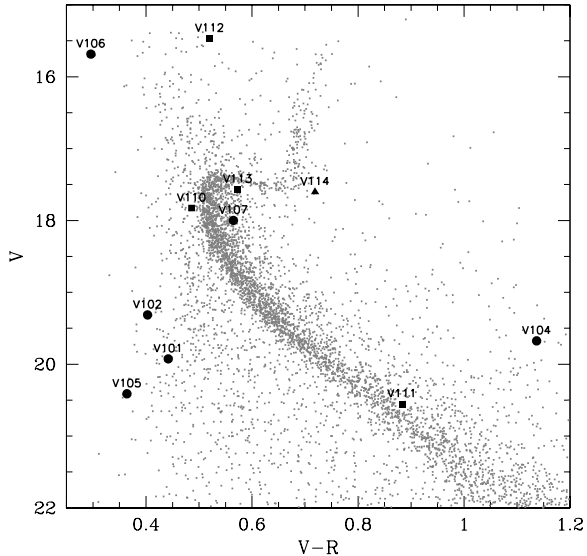}
\caption{$V/\vr$ color-magnitude diagram for Chip~3, centered on
NGC~6791. Newly discovered eclipsing binaries are plotted with
circles, other periodic variables with squares and the non-periodic
variable with a triangle.}
\label{fig:cmd}
\end{figure}

\begin{figure*}[t]
\epsscale{2.3}
\plottwo{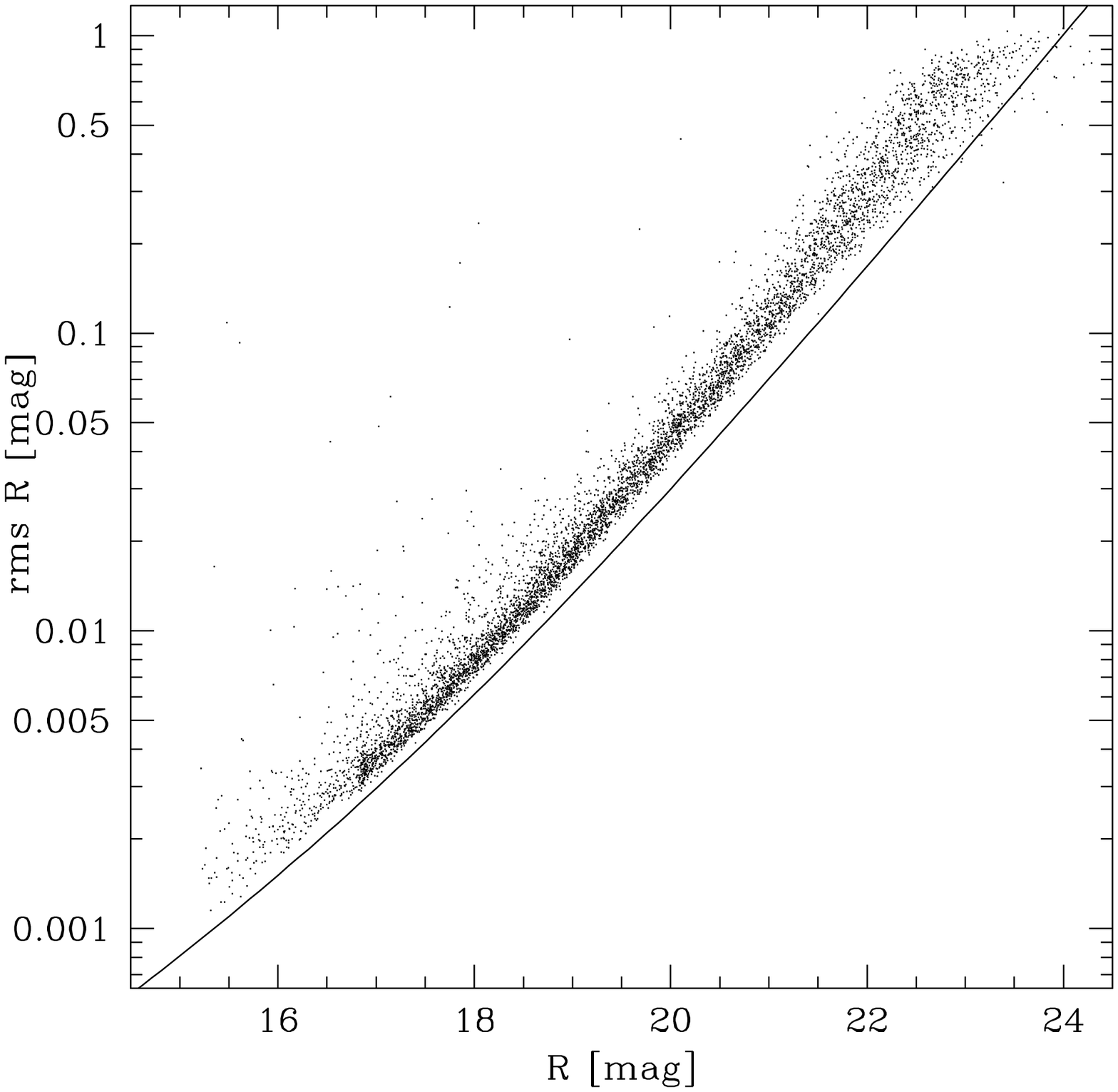}{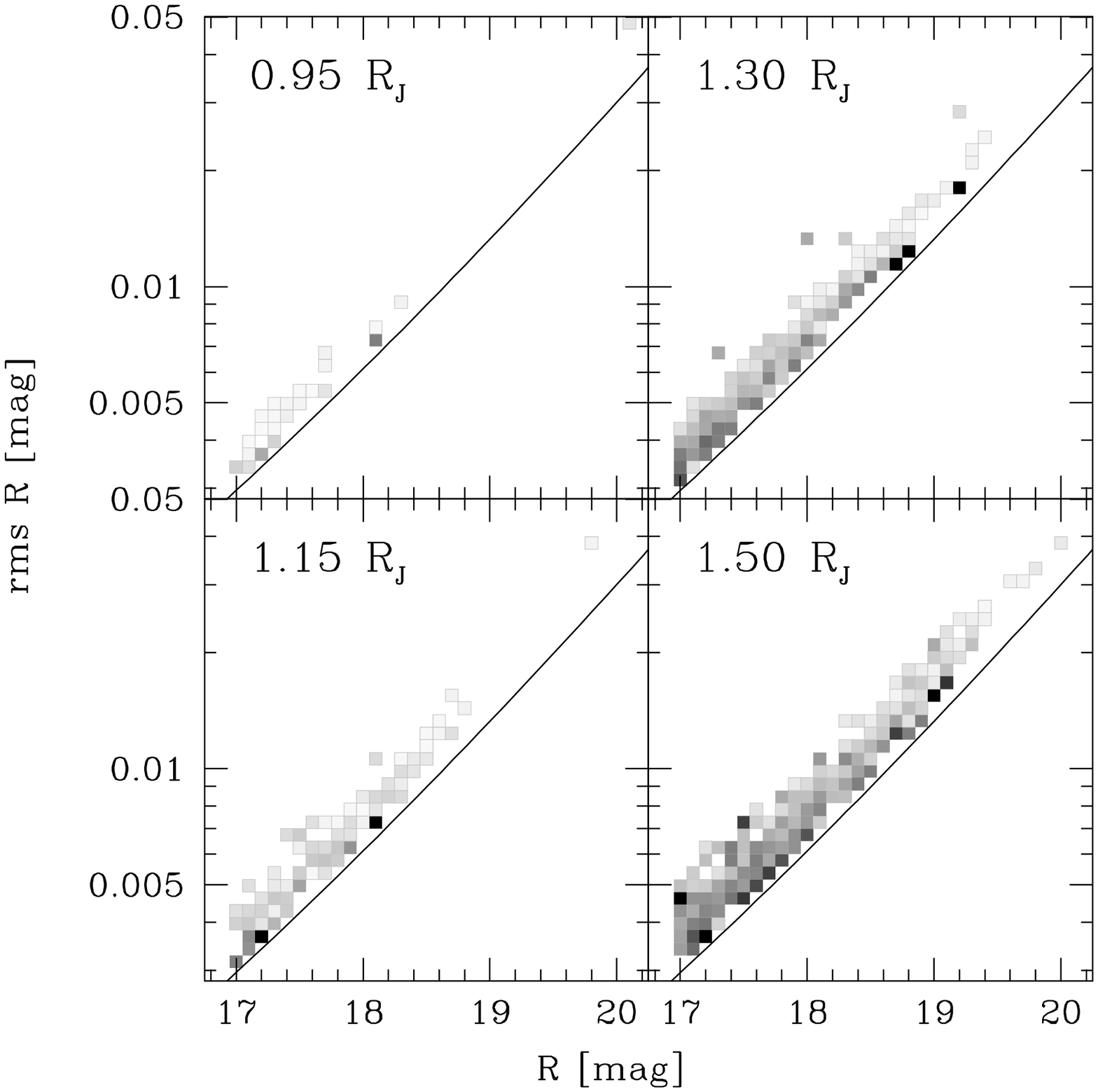}
\epsscale{1}
\caption{Left panel: The rms scatter of the $R$-band light curves for
stars on Chip~3 with at least 650 data points. The continuous curve
indicates the photometric precision limit due to Poisson noise of the
star and average sky brightness. Right panel: The detection efficiency
of 0.95, 1.15, 1.30 and 1.50 $R_J$ planets as a function of magnitude
and rms scatter (white: 0\%, black 100\%), determined in \S
\ref{sec:detef}.}
\label{fig:rms}
\end{figure*}

\subsection{{\it Calibration}}
The transformations of instrumental magnitudes to the standard system
were derived from observations of 15, 17, 15 and 17 stars on
Chips~1-4, respectively, in four Landolt (1992) standard fields,
collected on 2002 May 18.  Transformations in the following form were
adopted:
\begin{eqnarray*}
\label{eq:v}
v = V + a_1 + a_2 (V-R)+ a_3 (X-1.25)\\
\label{eq:vr}
v-r =  b_1 + b_2 (V-R)+ b_3 (X-1.25)\\
r = R + c_1 + c_2 (V-R)+ c_3 (X-1.25)
\end{eqnarray*}
where X is the airmass. Table~\ref{tab:cal} lists the coefficients
$a_i$, $b_i$, $c_i$ and the rms scatter between the observed and
calculated standard $VR$ magnitudes. 

These coefficients were used to calibrate the photometry from the
images of the cluster taken during the same night as the
standards. The magnitudes from the reference images were transformed
using the same color and extinction coefficients. The offsets were
determined relative to the calibrated photometry from cluster images
taken on the standard night. Figure~\ref{fig:cmd} shows the calibrated
V/V-R color-magnitude diagram (CMD) for the Chip~3 reference
image\footnote{The calibrated $VR$ photometry is available from the
authors via the anonymous ftp on cfa-ftp.harvard.edu, in the
/pub/bmochejs/PISCES directory.}.

A comparison of our $V$-band magnitudes with the photometry of
Stetson, Bruntt \& Grundahl (2003) reveals offsets of 0.048, 0.027,
0.047 and 0.009 mag in Chips 1-4, based on 20, 303, 3423 and 280 stars
above $V=20$, respectively. We also find an offset of 0.022 mag in $V$
between our Chip 3 and Mochejska et al.\ (2003) data. 

The $VR$ light curves were converted from differential flux to
instrumental magnitudes using the method described in Paper I.
Instead of Eq.~(1) from Paper I, we used the following relation to
compute the total flux corresponding to the $i$-th image, $c_i$:
\begin{equation}
c_i = c_{ref} - \Delta c_i
\end{equation}
where $\Delta c_i = c_{ref}-c_i$ is the flux on the $i$-th subtracted
image and $c_{ref}$ is the total flux on the reference image. This
method should yield more accurate results because it is based on the
reference image which has a higher S/N ratio than the template image
used previously. The instrumental magnitudes were transformed to the
standard system by adding offsets, computed individually for each
star, between the instrumental and calibrated reference image
magnitudes.

\subsection{{\it Astrometry}}
Equatorial coordinates were determined for the $R$-band reference
image star lists. The transformation from rectangular to equatorial
coordinates was derived using 964, 1012, 1476 and 951 transformation
stars from the USNO B-1 catalog (Monet et al.\ 2003) in Chips~1
through 4, respectively. The mean of the absolute value of the
deviation between the catalog and the computed coordinates for the
transformation stars was $0\farcs 13$ in right ascension and $0\farcs
11$ in declination.

\section{{\sc Search for Transiting Planets}}

\subsection{{\it Further Data Processing}}

We rejected 157 $R$-band epochs where less than 5000 stars were
detected on Chip~3 by DAOphot (Stetson 1987), and 25 other bad quality
images from three nights. This left us with 936 highest quality
$R$-band exposures with a median seeing of $2\farcs 1$.  We also
removed 8 $V$-band images, which left us with 318 exposures with a
median seeing of $2\farcs 3$.

In the light curves we noticed the presence of offsets between
different runs. This may be due to the periodic UV flooding of the CCD
camera, which alters its quantum efficiency as a function of
wavelength. To prevent the transit detection algorithm from mistaking
these changes in brightness for transits, we added offsets between the
runs, individually for each light curve, so that the median magnitude
was the same during each run. There were nine runs, each spanning from
44 to 187 data points.  Typical sizes of the offsets were 0.008 mag
for stars below $R=18$ and 0.018 mag for stars between $R=18$ and
19. As described in \S \ref{subsec:detef}, this procedure greatly
improves our detection efficiency.

The left panel of Fig.~\ref{fig:rms} shows the rms scatter of the
$R$-band light curves for stars on Chip~3 with at least 650 data
points. The continuous curve indicates the photometric precision limit
due to Poisson noise of the star and average sky brightness.  The
right panel shows the detection efficiency of 0.95, 1.15, 1.30 and
1.50 $R_J$ planets as a function of magnitude and rms scatter (white:
0\%, black 100\%), determined in \S \ref{sec:detef}.

\subsection{{\it Selection of Transiting Planet Candidates}}

For further analysis we selected stars with at least 650 out of 936
good epochs and light curve rms below 0.05 mag. Stars above the main
sequence turnoff ($R=17$) were rejected due to their large radii and,
hence, very small expected transit depths (below 0.4\%). This left us
with 3074 stars on Chip~3, and 2975 on Chips~1, 2 and 4 (825, 1091 and
1059 stars, respectively).

To select transiting planet candidates we used the box-fitting
least-squares (BLS) method (Kov{\' a}cs, Zucker, \& Mazeh 2002).
Adopting a cutoff of 6 in Signal Detection Efficiency (SDE) and 9 in
effective signal-to-noise ratio ($\alpha$), we selected 185 candidates
on Chip~3 and 39 on Chips~1, 2 and 4 (12, 16 and 11 candidates,
respectively). We found three candidates on Chip~3, which were rejected
as false detections upon closer examination. They had similar
coordinates on the image and their periods were all nearly exact
integral multiples of 0.9244 days. We have found 13 other stars within a
distance of 50 pixels whose periods were also such multiples. An
examination of the period distribution of all stars revealed significant
peaks at 4, 6 and 8 times 0.9244 days. We did not find any other good
transiting planet candidates.

\begin{figure}[t]
\plotone{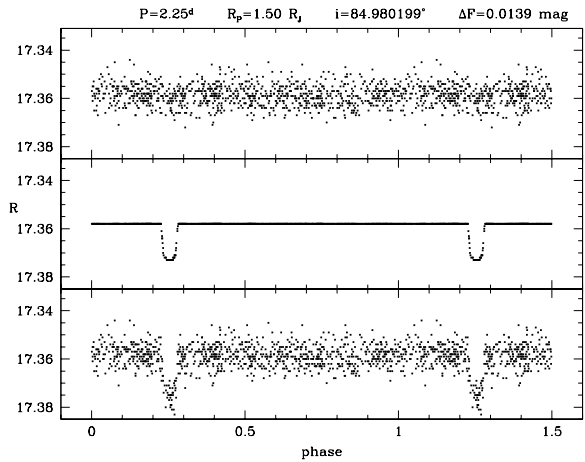}
\caption{The original, model and combined light curves (upper, middle
and lower panels, respectively) for a star with $R=18.01$ and a planet
with a period of 3.25 days, radius of 1.3 $R_J$ and inclination of
88$^{\circ}$.}
\label{fig:tr}
\end{figure}

\begin{figure*}[t]
\epsscale{2.3}
\plottwo{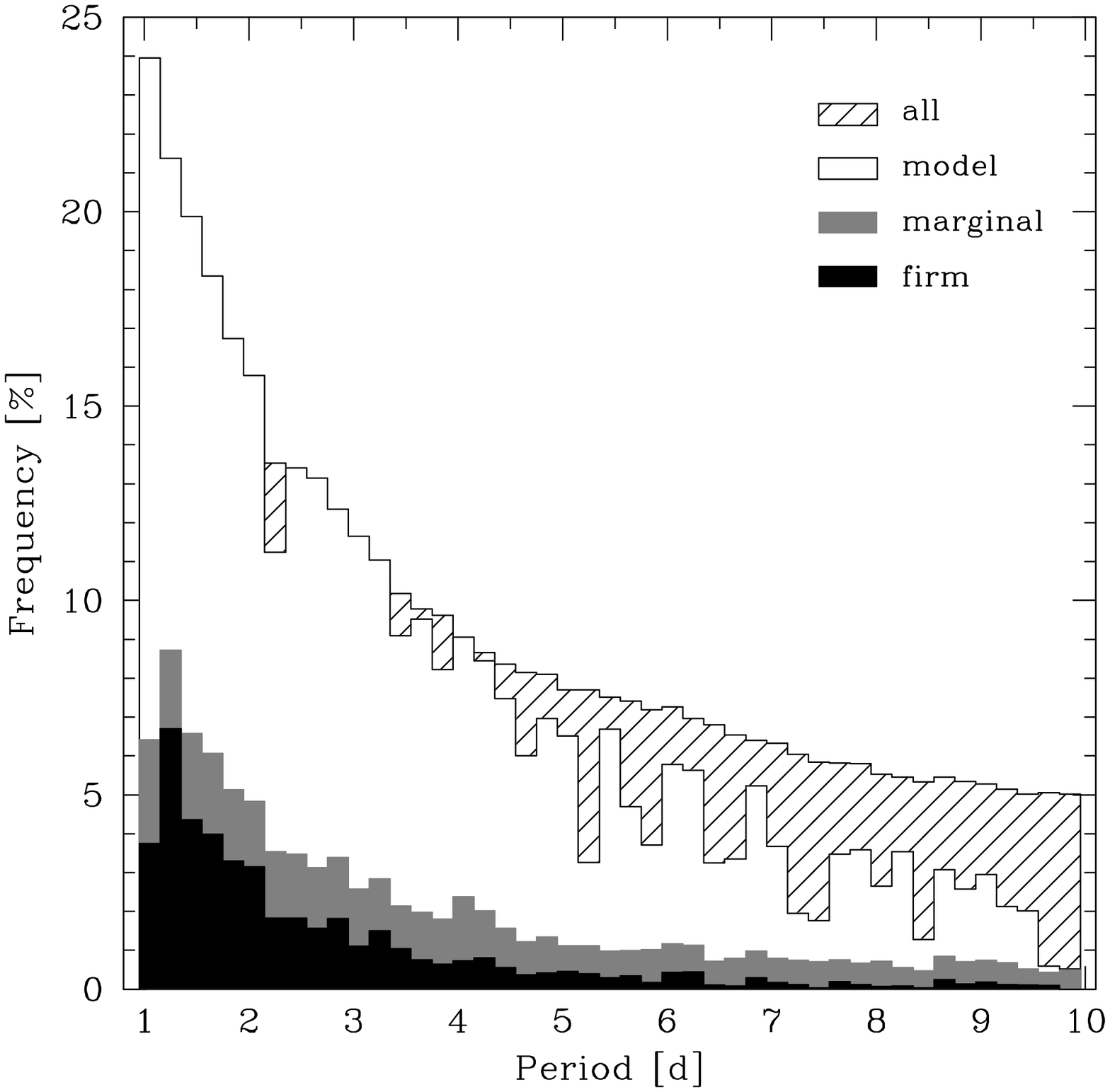}{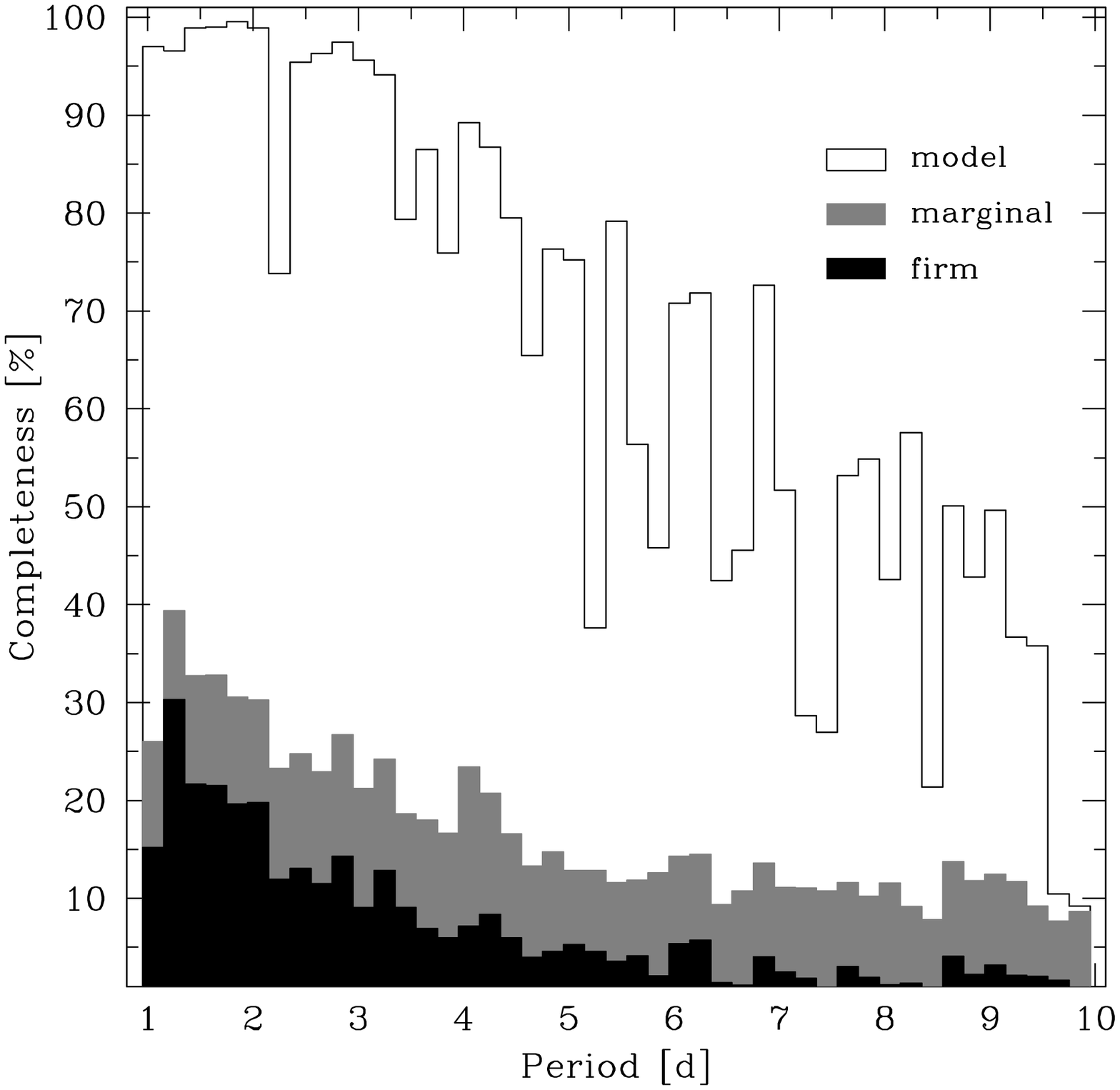}
\caption{Detection efficiency of transiting planets as a function of
their period, relative to planets with all inclinations (left panel)
and all transiting planets (right). Shown are the distributions for
all transiting planets (hashed histogram), detections in the
model light curves (open) and marginal (gray) and firm (solid)
detections in the combined light curves. }
\label{fig:comp}
\end{figure*}

\section{{\sc Estimate of the Number of Expected Detections}}

The number of transiting planets we should expect to find, $N_P$, can
be derived from the following equation:
\begin{equation}
N_P = N_* f_P D
\end{equation}
where $N_*$ is the number of stars with sufficient photometric
precision, $f_P$ is the frequency of planets within the investigated
period range and $D$ is the detection efficiency, which accounts for
random inclinations. In \S\S \ref{sec:freq}, \ref{sec:nstar} and
\ref{sec:detef} we determine $f_P$, $N_*$ and $D$.

\subsection{{\it Planet Frequency}}
\label{sec:freq}

The frequency of planets is known to increase with the host star's
metallicity. From Figure~7 in Santos et al.\ (2004), the frequency of
planets for stars with [Fe/H] = +0.3 $-$ +0.4 dex is $\sim$28\%, and
it drops to $\sim$2.5\% for metallicities below [Fe/H] = +0.1 dex.

The percentage of planets with periods below 10 days is $14.6\%$ in
the Santos et al.\ (2004) sample. As of 16 November 2004, the
corresponding fractions for the planet lists on
exoplanets.org\footnote{http://exoplanets.org/planet\_table.txt} and
The Extrasolar Planets
Encyclopaedia\footnote{http://www.obspm.fr/encycl/cat1.html} were
$15.3\%$ and $15.8\%$ (excluding planets detected via transits). In
further analysis, we adopt the value of $15\%$ as the fraction of
planets with periods below 10 days.

Combining these two numbers yields $f_P$=4.2\% for the high
metallicity cluster stars and 0.375\% for field stars. Please note
that the latter estimate is considerably lower than the commonly
adopted frequency of 1\%.

\subsection{\it The Number of Cluster and Field Stars}
\label{sec:nstar}

Most of the cluster is contained on Chip~3 but its main sequence (MS)
is also discernible on Chips~1, 2 and 4. To obtain a rough estimate of
the number of stars belonging to the cluster, we determined the MS
fiducial line and counted as members all stars within 0.06 mag of it
in \vr, on all four chips. This gives 246, 381, 2201 and 350
``cluster'' stars and 577, 710, 852 and 706 ``field'' stars on
Chips~1-4, respectively. Twenty six stars did not have V-band data,
and we assumed that they belong to the field. There are a total of
3178 ``cluster'' and 2871 ``field'' stars. There are more ``field''
stars on Chip~3 than on the other chips, which means that some of them
belong to the cluster and our color cutoff is not too liberal. On the
other hand, a small fraction of the ``cluster'' stars are field stars,
so these two biases should cancel out to some extent.

\begin{figure*}[htb]
\plottwo{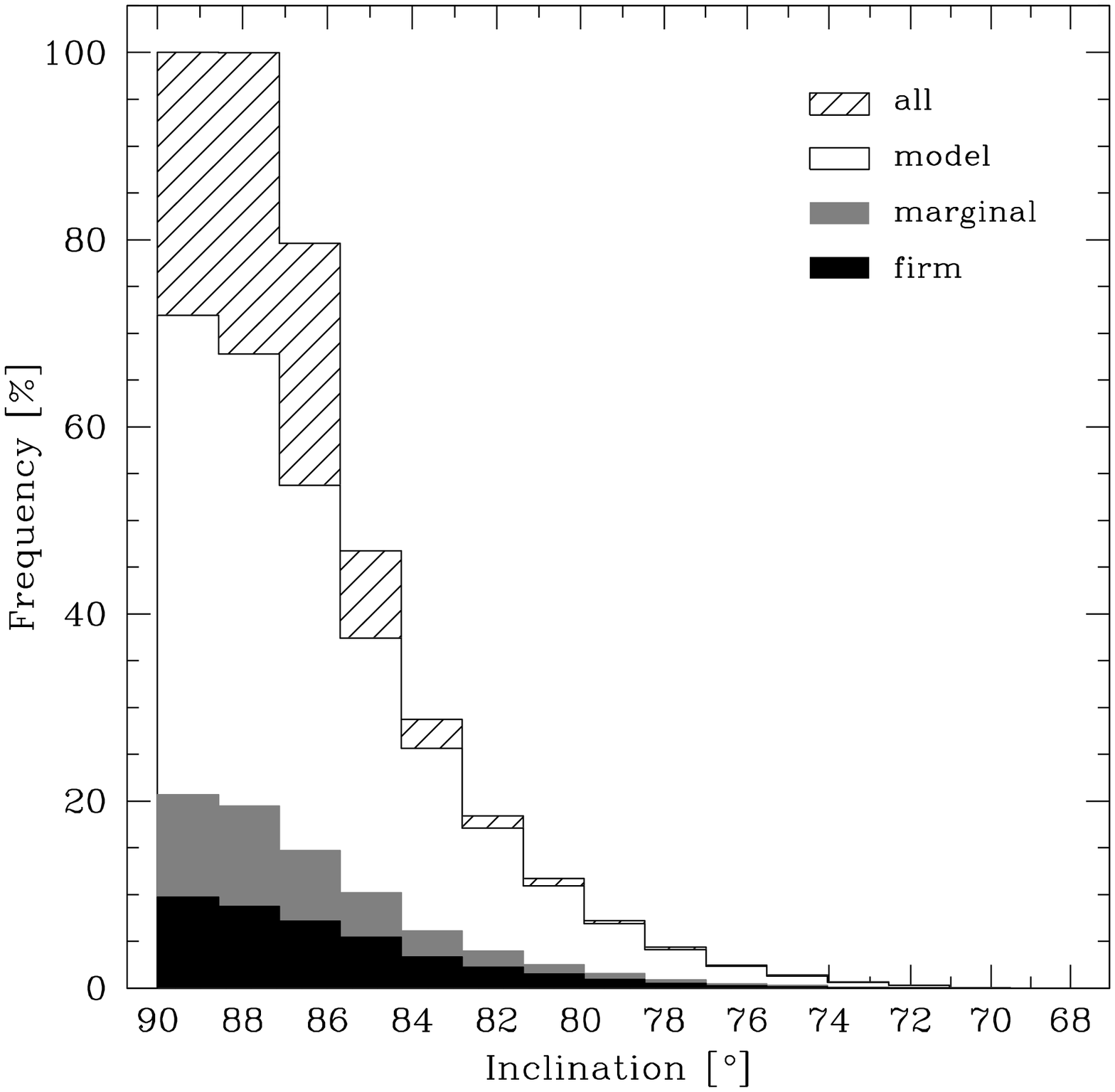}{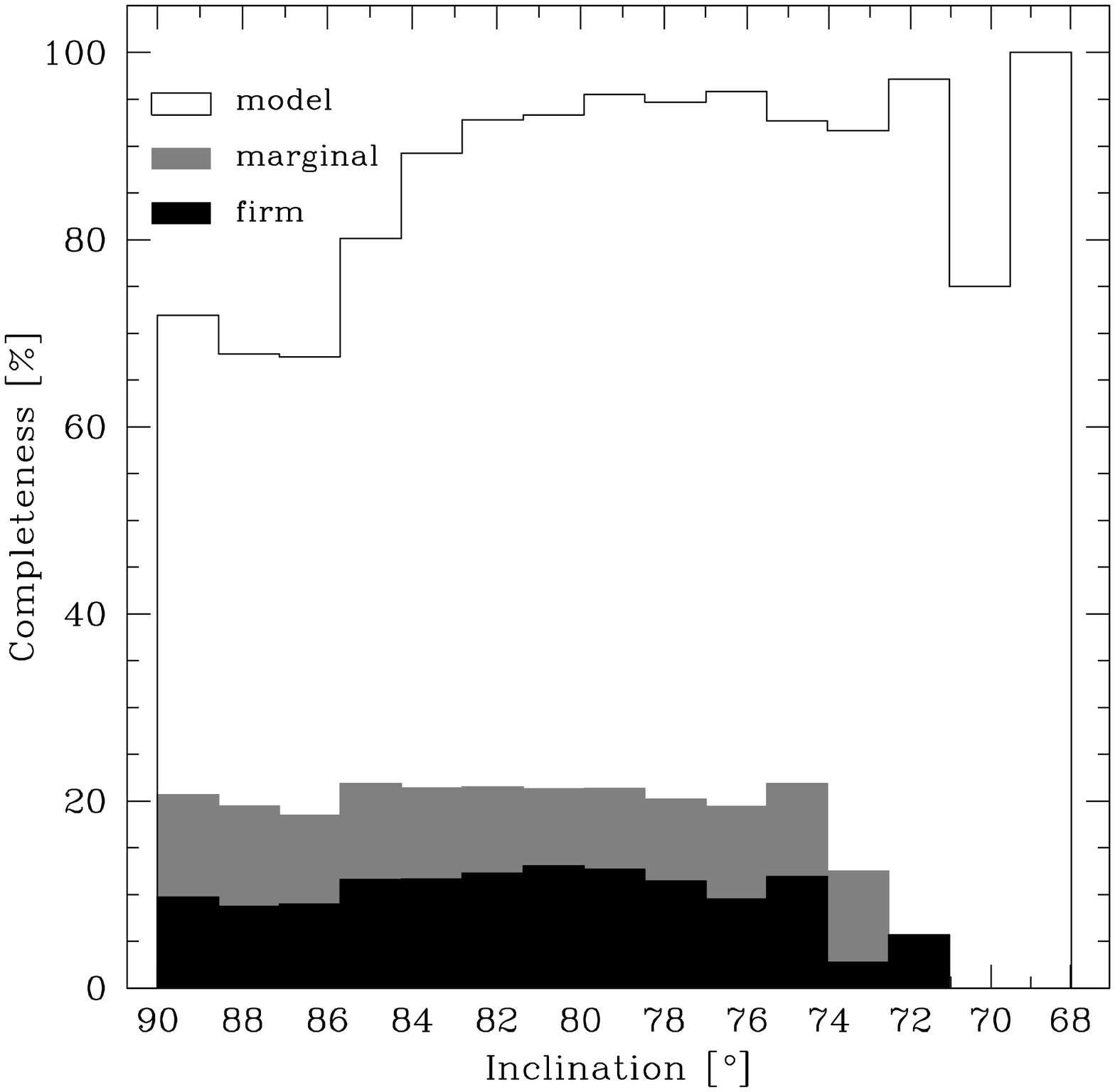}
\caption{Detection efficiency of planetary transits as a function of
their inclination, relative to planets with all inclinations (left panel)
and all transiting planets (right). }
\label{fig:i}
\end{figure*}

\begin{figure*}[htb]
\plottwo{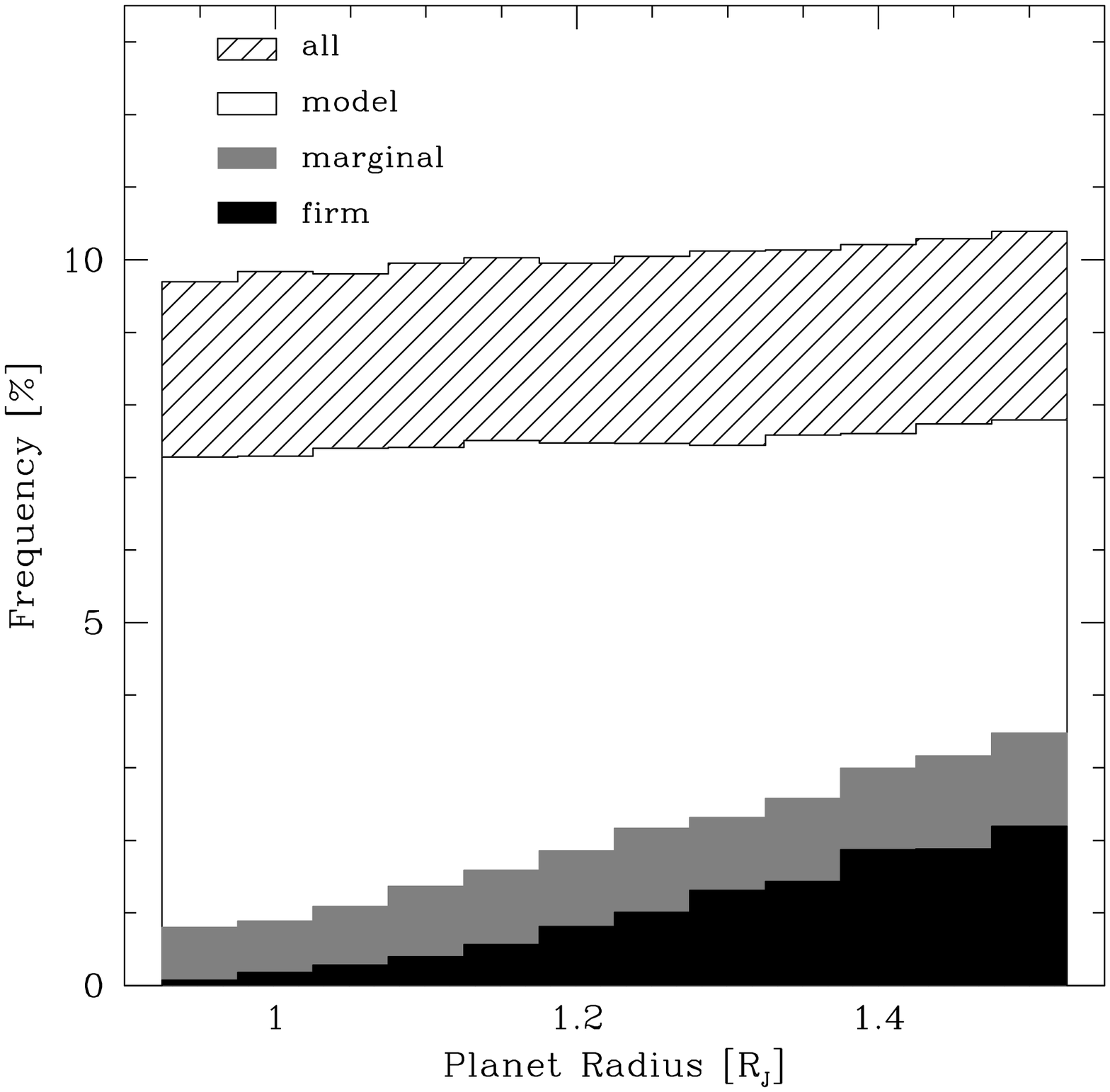}{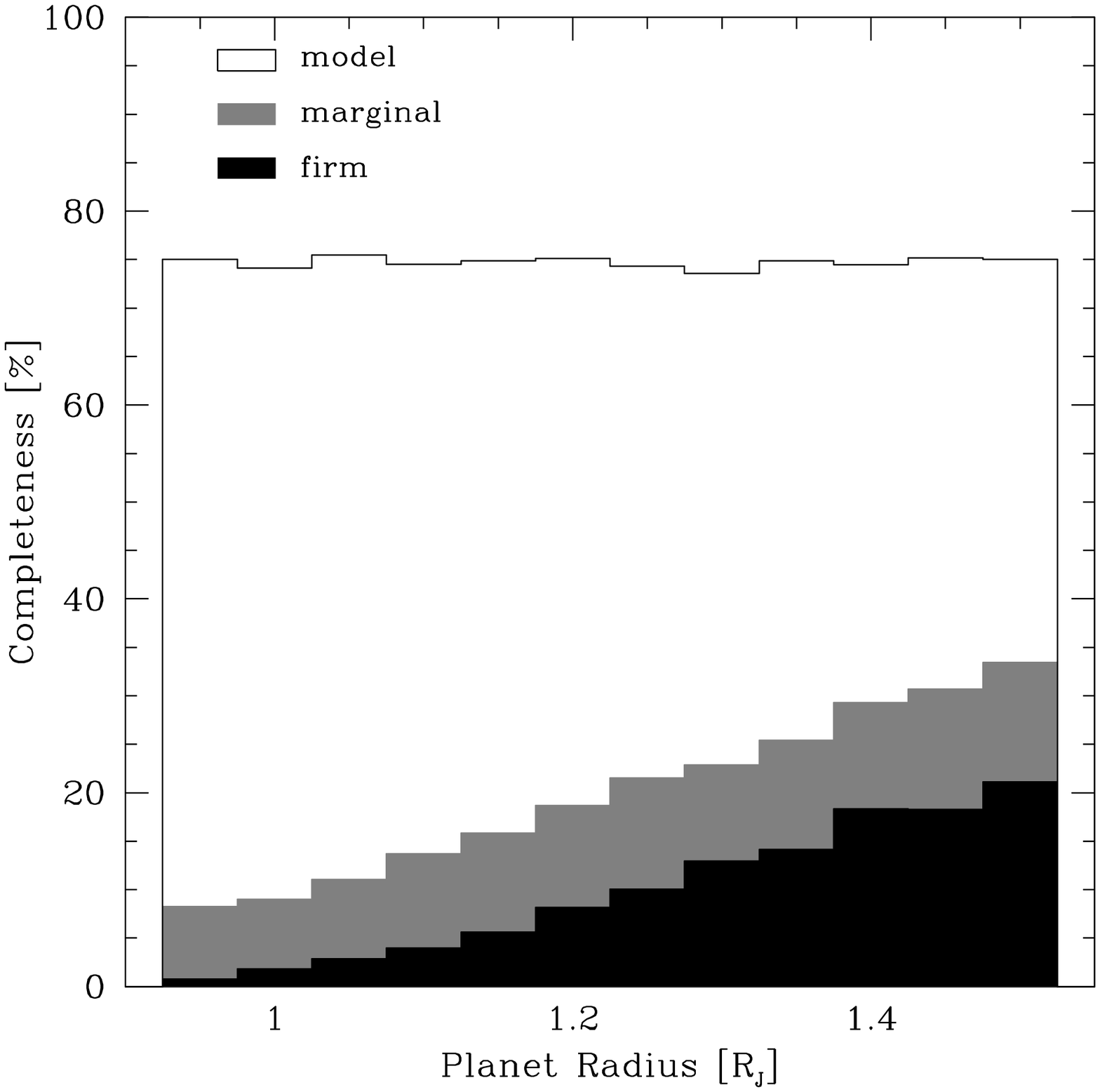}
\caption{Detection efficiency of planetary transits as a function of
their radius, relative to planets with all inclinations (left panel)
and all transiting planets (right). }
\label{fig:rp}
\end{figure*}

\begin{figure*}[htb]
\plottwo{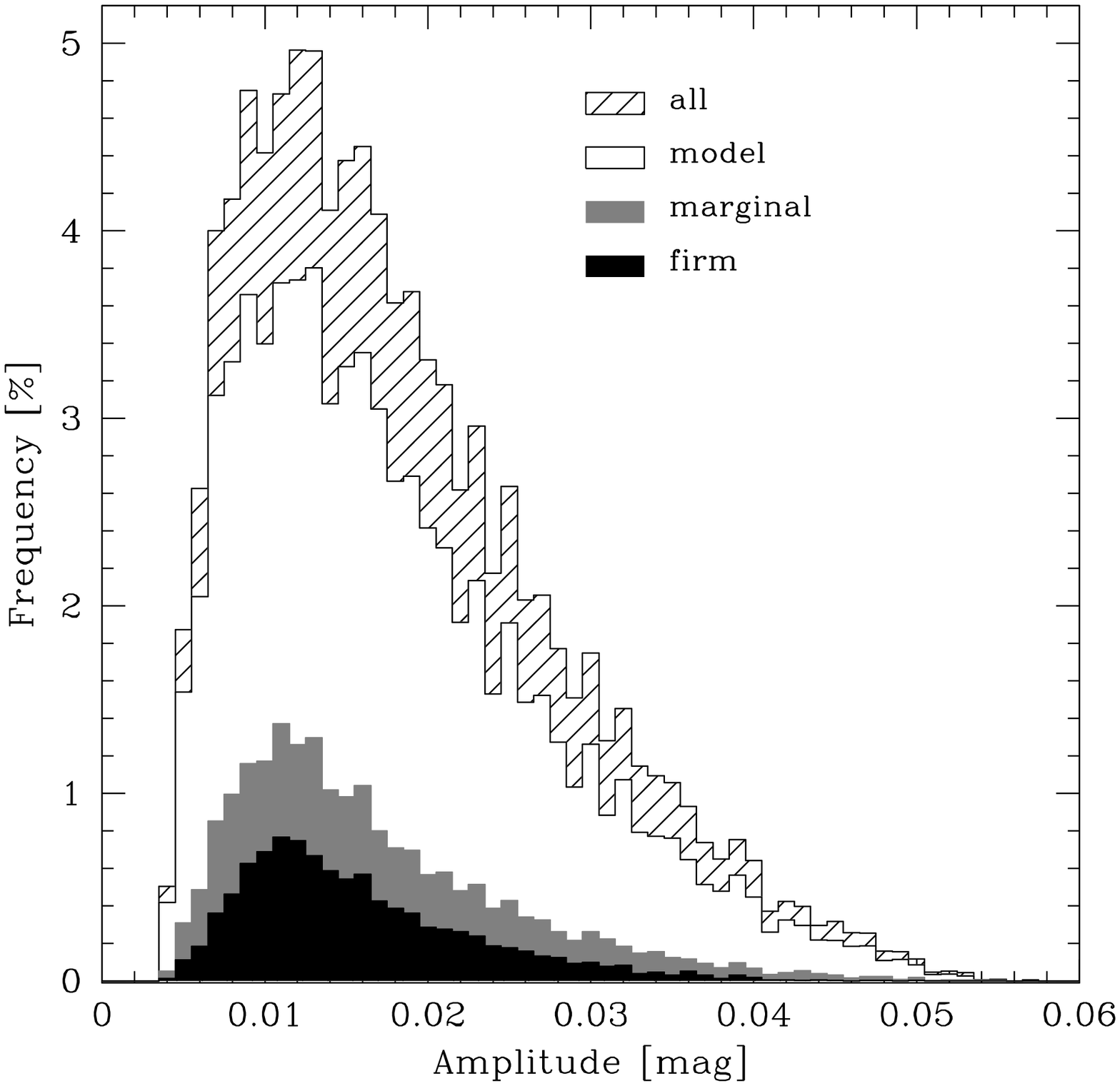}{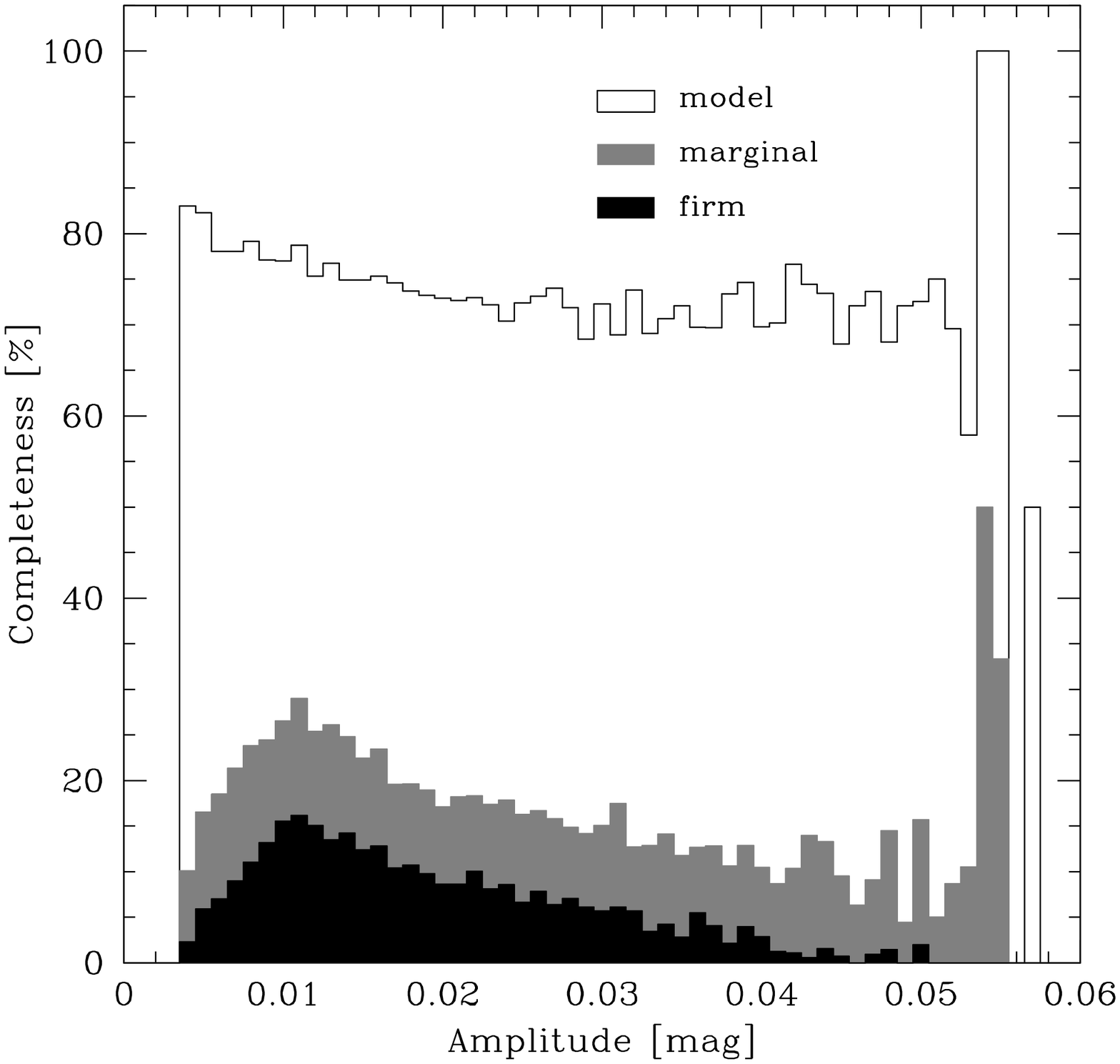}
\caption{Detection efficiency of planetary transits as a function of
their amplitude, relative to planets with all inclinations (left panel)
and all transiting planets (right). }
\label{fig:amp}
\end{figure*}

\subsection{{\it Detection Efficiency}}
\label{sec:detef}
In order to characterize our detection efficiency, we inserted model
transits into the observed light curves, and tried to recover them
using the BLS method.

\subsubsection{{\it Model Transit Light Curves}}
The model transit light curves were defined by five parameters: the
transit depth, $\Delta$F, total transit duration, $t_T$, transit
duration between ingress and egress, $t_F$ (the ``flat'' part of the
transit), the period of the planet, $P$ and the limb darkening
coefficient, $u$.

The first three parameters were computed from equations (1), (15) and
(16) in Seager \& Mall{\' e}n-Ornelas (2003):

{\footnotesize
\begin{eqnarray}
\label{eq:df}
\Delta F & = & \left(\frac{R_p}{R_*}\right)^2\\
\label{eq:tF_tT}
\left(\frac{t_F}{t_T}\right)^2 & = & \frac{ \left( 1 -
\frac{R_p}{R_*}\right)^2 - \left(\frac{a}{R_*} \cos i\right)^2}
{\left( 1 + \frac{R_p}{R_*}\right)^2 - \left(\frac{a}{R_*} \cos
i\right)^2}\\
\label{eq:tT}
t_T & = & \frac{P R_*}{\pi a}\sqrt{\left(1 + \frac{R_P}{R_*}\right)^2
- \left(\frac{a}{R_*} \cos i\right)^2}
\end{eqnarray}}

Equations (\ref{eq:tF_tT}) and (\ref{eq:tT}) are valid for $t_T\pi P
\ll 1$. The radius of the planetary orbit, $a$, can be derived from
the star's mass, $M_*$, and Kepler's third law, with the planet's mass
$M_p \ll M_*$:

{\footnotesize
\begin{equation}
\label{eq:a}
 a = \left[ \frac{P^2 G M_*} {4 \pi^2}\right]^{1/3}
\end{equation}}

The radius and mass of the star, $R_*$ and $M_*$, were interpolated,
as a function of absolute $R$-band magnitude, $M_R$, from the highest
metallicity (Z = 0.03) 7.943 Gyr isochrone of Girardi et al.\
(2000). A distance modulus $(m-M)_R=13.36$ mag was used to bring the
observed $R$-band magnitudes to the absolute magnitude scale (Chaboyer
et al.\ 1999).

The effects of limb darkening were simulated using the linear
approximation first introduced by Milne (1921):

{\footnotesize
\begin{equation}
\label{eq:mu}
 I(\mu) = (1 - u (1 - \mu))
\end{equation}}

where $u$ is the limb darkening coefficient, $\mu=cos(\theta)$,
$\theta$ is the angle between the line of sight and the emergent flux,
and $I(1)$ is the intensity at the center of the disk. We used the
grid of $R$-band limb darkening coefficients, given as a function of
gravity, $G$, and temperature, $T_{eff}$, by Claret, Diaz-Cordoves \&
Gimenez (1995). For each star, its $G$ and $T_{eff}$ were determined
from Girardi et al.\ (2000) isochrones and $u$ was interpolated from
four closest points in $G$ and $T_{eff}$ in the Claret et al.\ (1995)
grid.

In addition to $P$, the equations contain two other free parameters:
the planet radius, $R_P$ and the inclination of the orbit, $i$. A
fourth parameter which affects the detectability of a planet is the
epoch of the transits, $T_0$.

\subsection{\it Test Procedure}

We investigated the range of parameters specified in Table
\ref{tab:pars}, where $P$ is expressed in days, $R_P$ in Jupiter radii
($R_J$), $T_0$ as a fraction of period. We examined the range of
periods from 1.05 to 9.85 days and planet radii from 0.95 to 1.5
$R_J$, with a resolution of $0.2$ days and $0.05$ $R_J$,
respectively. For $T_0$ we used an increment of 5\% of the period, and
a 0.025 increment in $\cos i$. The total number of combinations is
432000.

For each combination of parameters, a random star was chosen without
replacement from the sample of 3074 stars on Chip~3. When the sample
was exhausted, it was reset to the original list.  The ``observables''
$\Delta$F, $t_T$ and $t_F$ were computed and when $t_T \geq 0.5^h$ two
light curves were generated: the {\bf model} transit light curve, and
the observed light curve combined with the model (hereafter referred
to as the {\bf combined} light curve).  Figure~\ref{fig:tr} shows the
original, model and combined light curves (upper, middle and lower
panels, respectively) for a star with $R=17.36$ and a planet with a
period of 2.25 days, radius of 1.5 $R_J$ and inclination of
85$^{\circ}$. The amplitude of the transit is 0.0139 mag, and the mass
and radius of the star, taken from the models, are 1.03 $M_{\odot}$
and 1.28 $R_{\odot}$.

To assess the impact of the procedure to correct for offsets between
the runs on our detection efficiency, we investigated three cases,
where the correction was applied:
\begin{itemize}
\vspace{-0.2cm}
\item [A.]{after inserting transits,}
\vspace{-0.2cm}
\item [B.]{before inserting transits,}
\vspace{-0.2cm}
\item [C.]{was not applied at all.}
\vspace{-0.2cm}
\end{itemize}
Case (B) will give us the detection efficiency if our data did not
need to be corrected, and case (C) if we did not apply the
corrections. Case (A) will give us our actual detection efficiency,
and its comparison with cases (B) and (C) will show how it is affected
by the applied correction procedure.

This required us to run two sets of simulations: on the original
(cases A and C) and corrected (case B) light curves. In both
simulations the same list of parameter and star combinations was used.

\subsection{\it Detection Criteria}

An examination of the frequency of recovered periods, relative to the
input period, revealed that only the peaks at 1, 2 and $\frac{1}{2}$
$P_{inp}$ are distinct. Other aliases blend in with the background of
the incorrectly recovered periods, so we have disregarded them.

A transit was flagged as detected if:
\begin{enumerate}
\vspace{-0.2cm}
\item {The period recovered by BLS was within 2\% of the input period
$P_{inp}$, 2 $P_{inp}$ or $\frac{1}{2}$ $P_{inp}$,}
\vspace{-0.2cm}
\item {The BLS statistics were above the following thresholds: $SDE >
6$, $\alpha > 9$.}
\vspace{-0.2cm}
\end{enumerate}
These detections will be referred to hereafter as {\it firm}.
Detections where only condition (1) was fulfilled will be called {\it
marginal}.

\subsection{\it Detection Efficiency}
\label{subsec:detef}
The results of the tests are summarized in Table~\ref{tab:art}, which
lists the test type (A-C), the number and percentage of transits with
$t_T \geq 0.5^h$ (out of the 432000 possible parameter combinations),
and the numbers and percentages (relative to the total number of
transits in column 2) of transits detected in the model light curves,
and of marginal and firm detections in the combined light curves.

Figures~\ref{fig:comp}-\ref{fig:amp} show the dependence of the
detection efficiency on period, inclination, planet radius and transit
amplitude. The hashed, open, gray and solid histograms denote
distributions for all transiting planets, planets detected in the
model light curves, and marginal and firm detections in the combined
light curves, respectively. Left panels show the frequency of transits
and transit detections relative to planets with all inclinations.
Right panels show the detection completeness normalized to all
transiting planets (plotted as hashed histograms in left panels).

The tests show that 10\% of planets with periods 1-10 days will
transit their parent stars. This frequency drops from $\sim$24\% at
$P=1^d$ to $\sim$5\% at $P=10^d$. All planets with inclinations
$87-90^\circ$ transit their host stars, and this fraction drops to
$\sim$80\% for $i=86^\circ$ and $\sim$5\% for $i=78^\circ$. The
frequency of transits increases very weakly with planet radius. The
distribution of transit amplitudes has a wide peak stretching from
0.6\% to 2\%, centered on $\sim$1.3\%.

The percentage of detections for the model light curves illustrates
the limitation imposed on our detection efficiency by the temporal
coverage alone. Due to incomplete time sampling, we are restricted to
75\% of all planets with periods between 1 and 10 days. For periods
below 4 days, our temporal coverage is sufficient to detect $\sim$90\%
of all transiting planets, and drops to $\sim$50\% at $P=9$ days. The
detection completeness increases with decreasing inclination because
at lower $i$ only short period planets can transit their host
stars. It does not depend on the planet radius, and it decreases with
increasing transit amplitude.

The source of the dependence of the detection completeness on transit
amplitude is not as straightforward as for the other correlations. The
amplitude depends on the radii of the star and planet. Since the
detection completeness was found to be largely independent of the
planet radius, the observed trend must stem from its dependence on the
host star's radius, which is a function of its magnitude. Such a
correlation is indeed observed, with completeness increasing for
brighter stars (not shown here). The link between the temporal
coverage and magnitude comes from the observed increase in the number
of points in the light curve with decreasing magnitude.

For cases A, B and C, we {\it marginally} detect 20\%, 21\% and 13\%
of all transiting planets, and {\it firmly} detect 10\%, 11\% and
4.6\%, respectively. Transiting planets with firm detections
constitute 83\%, 84\% and 64\% of all stars with $SDE > 6$ and $\alpha
> 9$. Adding offsets between runs (case A) decreases the number of
firm detections by 7\%, compared to the desired case, where no offsets
would be required (case B). If the offsets were not corrected (case
C), we would detect only 46\% of the transiting planets detected in
case A.

The detection completeness for firm detections peaks at 20\% for
periods $1-2^d$ and decreases with period more steeply than model
detections. It does not show a marked dependence on inclination, and
strongly increases with increasing planet radius, from below 2\% at 1
$R_J$ to over 20\% at 1.5 $R_J$. This is also apparent in the right
panel of Fig.~\ref{fig:rms}, which shows the detection efficiency of
0.95, 1.15, 1.30 and 1.50 $R_J$ planets as a function of magnitude
and rms scatter (white: 0\%, black 100\%).

The detection efficiency peaks at an amplitude of $\sim$1\%, due to
the most favorable ratio between the transit amplitude and photometric
accuracy for this amplitude/magnitude range.

The efficiency of {\it firm} transiting planet detections, relative to
planets with all orbital inclinations, $D$, is $4323/432000=1.0\%$.

\begin{figure*}[!htb]
\epsscale{2}
\plotone{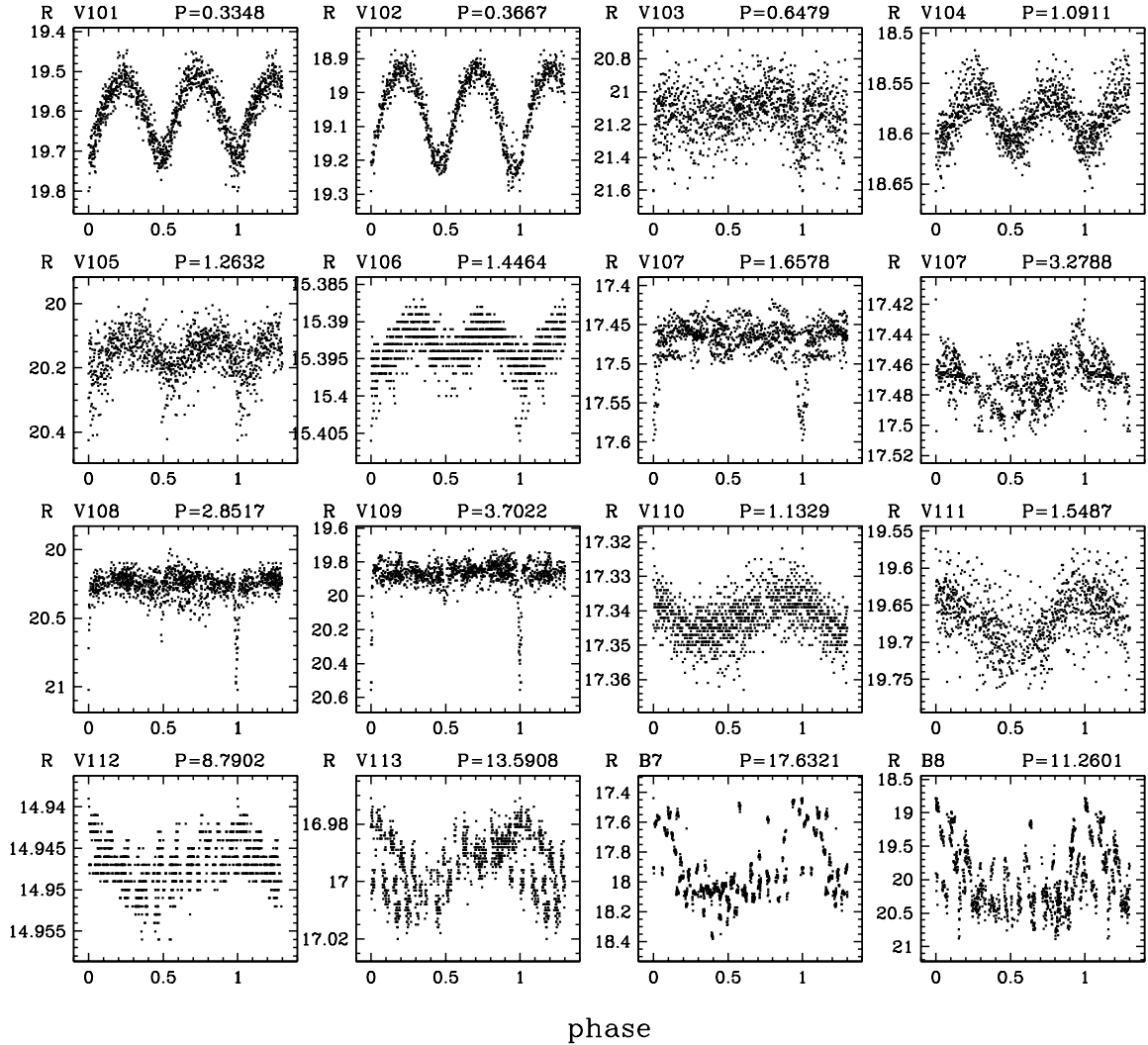}
\caption{The $R$-band light curves of nine new eclipsing binaries,
four new other periodic variables, and cataclysmic variables B7 and
B8. V107 is phased with two detected periods.}
\label{lc:ecl}
\end{figure*}

\begin{figure*}[!ht]
\plotone{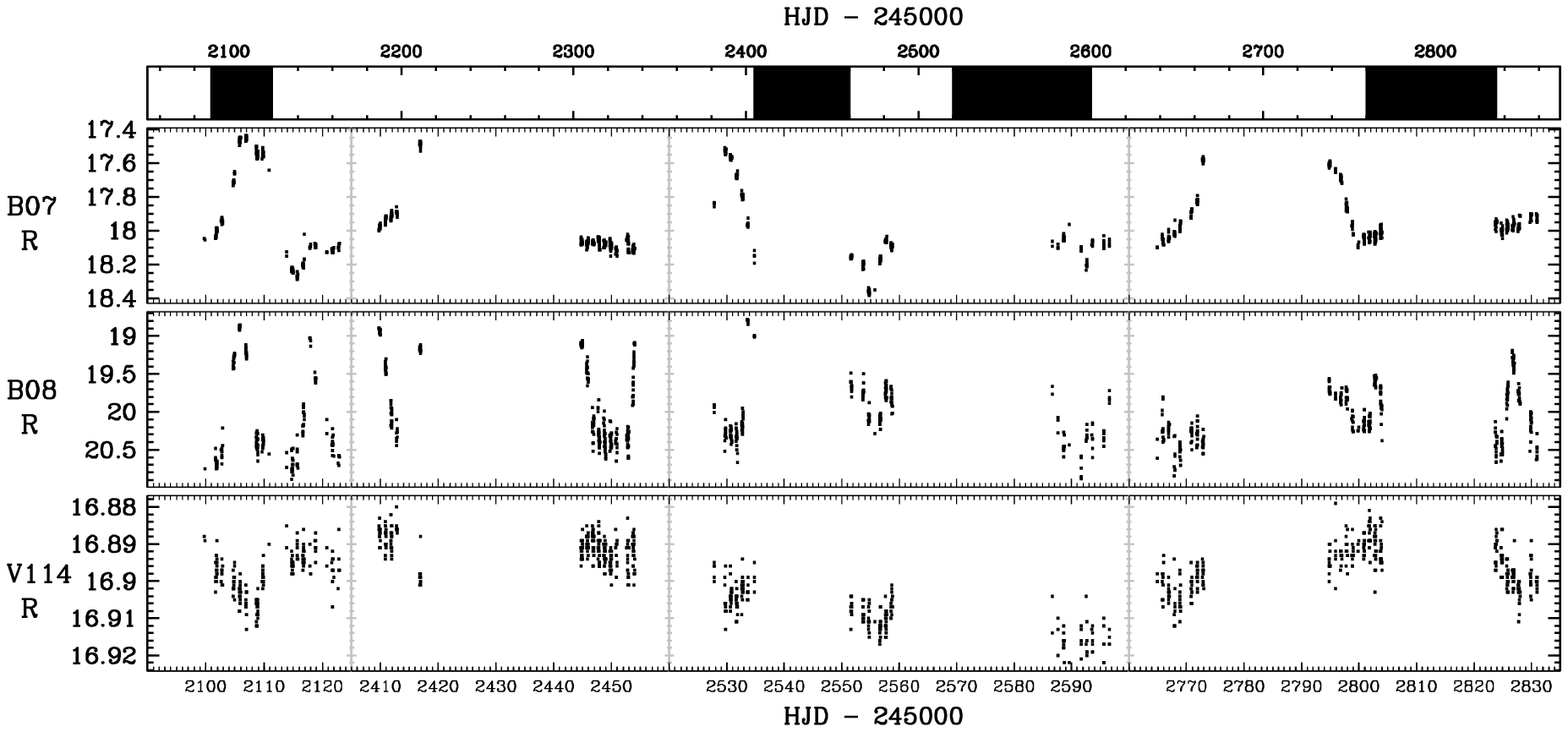}
\caption{The $R$-band light curves of the cataclysmic variables B7 and
B8 and the new variable V114. The top window illustrates the
distribution in time of the four sub-windows plotted for the variables.}
\label{lc:misc}
\end{figure*}

\subsection{\it Number of Transiting Planets Expected}

In \S\S~\ref{sec:freq}-\ref{sec:detef} we determined the planet
frequency $f_P$ to be $4.2\%$ for cluster stars and $0.375\%$ for
field stars, the number of stars in the cluster and field as 3178 and
2871 and our detection efficiency $D$ to be $1.0\%$. We should thus
expect 1.34 transiting planets in the cluster and an additional 0.11
of a planet among field stars.

\subsection{\it Discussion}

Figure~\ref{fig:comp} demonstrates that our temporal coverage is not the
limiting factor. To increase the number of expected planets it would be
necessary to improve the photometric precision.  The weather and seeing
conditions turned out to be inferior to what we were expecting. A better
quality CCD and a telescope with a larger diameter and/or better
observing conditions would be required to improve the chances for a
successful transiting planet search in NGC~6791.

The estimate of 1.45 expected transiting planets is not high enough to
enable us to draw any conclusions from the fact that we have not
detected any such events.

The precision of this estimate is largely limited by the uncertainty
in one of our basic assumptions -- the distribution of planetary
radii. This distribution is not precisely known, and changing it will
have a marked effect on the final result. Adopting a distribution of
planetary radii from 1.0 to 1.35 $R_J$, corresponding to the radius range
spanned by the six known transiting planets (Konacki et al.\ 2004)
would lower D from 1.0\% to 0.7\%. This translates to 1.08 detections,
compared to 1.45 with the original radius distribution -- a 26\%
decrease.

In Paper I we made the assumption that the planetary radii would span
the range 1-3 $R_J$, based on the radius of $1.347$ $R_J$ for the only
known transiting planet at the time, HD 209458b (Brown et al.\
2001). A simulation for planets in the radius range 1.5-3.0 $R_J$
shows that 11\% of them transit their parent stars, 75\% are detected
in the model light curves, and 56\% and 45\% are marginal and firm
detections in the combined light curves. Assuming that planet radii
are distributed evenly between 1 and 3 $R_J$ would give the percentage
of firm detections of 37\% and detection efficiency $D=4.0\%$, which
translates into 5.33 expected detections in the cluster and 0.43 in
the field. Our lack of detections does not favor such large planetary
radii, in agreement with observations (Fig.\ 5 in Konacki et al.\
2004) and current models (Bodenheimer, Laughlin \& Lin 2003; Burrows
et al.\ 2004; Chabrier et al.\ 2004, Kornet et al.\ 2005).

\begin{figure*}[!ht]
\plotone{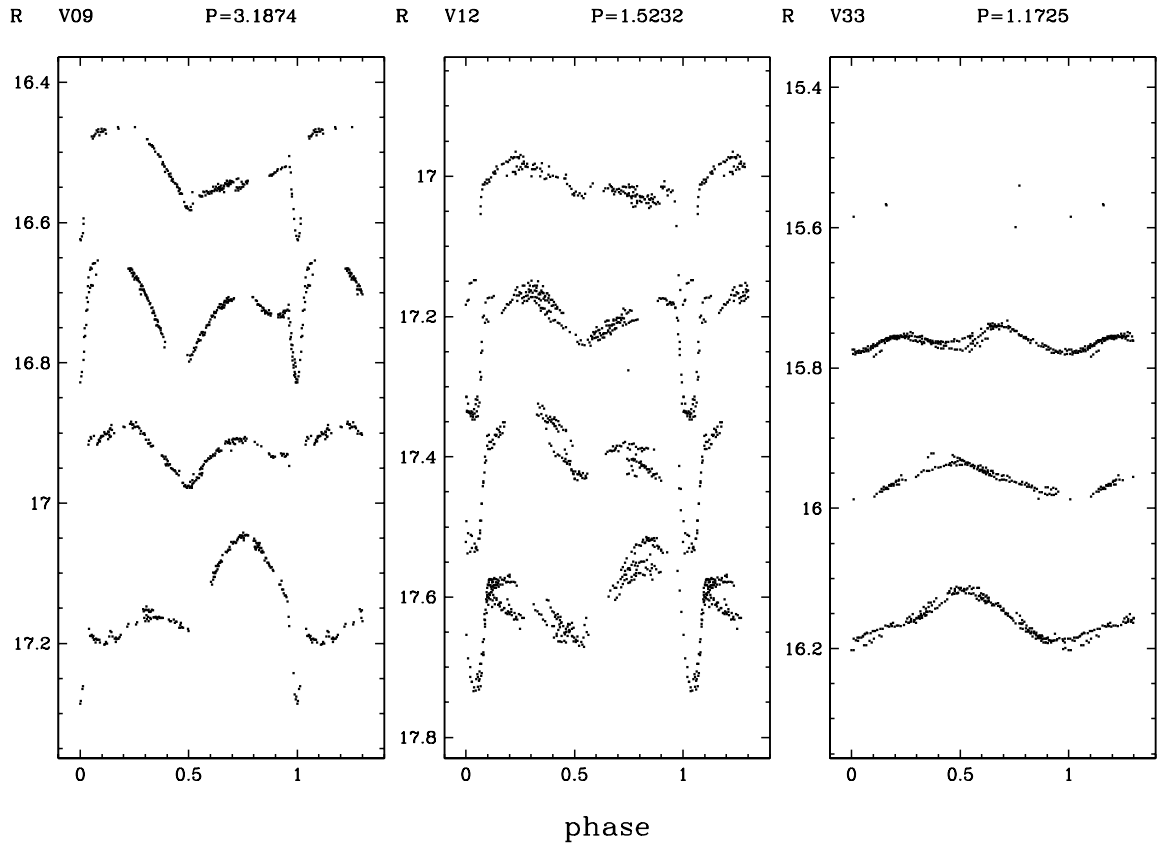}
\caption{The $R$-band light curves of the eclipsing binaries V9, V12
and V33.}
\label{lc:rscvn}
\end{figure*}

\section{\sc Previously Reported Candidates}
We examined the light curves of the transiting planet candidates
reported by Bruntt et al.\ (2003). None of them were found to exhibit
convincing periodic transits or eclipses. Even with our lower
photometric accuracy, in most cases we should have detected a
periodicity, if we had observed several transits. The rms of our
R-band light curves is $0.009$ for T10 and $0.003-0.006$ for the
remaining candidates.

\section{\sc Variable Stars}
We also extracted the light curves of known variable stars and
searched for new ones by running BLS in the period range 0.1-10 days.
In Tables \ref{tab:ecl}-\ref{tab:misc} we list their revised
parameters. We note, for the record, that variables V85, V88 and V96,
reported as new discoveries by Bruntt et al.\ (2003), are the same as
V76, V77 and V56 reported earlier by Kaluzny (2003) and in Paper~I.
We have reclassified V40 and V41 as eclipsing binaries.

We have discovered 14 new variables: 1, 4, 7 and 2 on Chips~1-4,
respectively. Their parameters are listed in Tables~\ref{tab:ecl} and
\ref{tab:misc} and their light curves are shown in
Figures~\ref{lc:ecl} and \ref{lc:misc}.\footnote{The $VR$ band
photometry and finding charts for all variables are available from the
authors via the anonymous ftp on cfa-ftp.harvard.edu, in the
/pub/bmochejs/PISCES directory.} They are also plotted on the CMD in
Fig.~\ref{fig:cmd}. Variables V101-V109 are eclipsing binaries: V101,
V102, V104-V106 are W UMa type contact systems, V103, V107-V109 are
detached or semi-detached binaries. In addition to eclipses, V107
displays out of eclipse variability with a period almost exactly twice
as long as the orbital period. In Fig.~\ref{lc:ecl} it is shown phased
with both periods. For clarity, points in the eclipse have been
removed from the light curve phased with the longer period. The other
periodic variables, V110-V113, are most likely spotted stars. The
shape of the light curve of V113 varies noticeably with time.

The non-periodic variable V114 is located slightly redward of the base
of the red giant branch. If it belongs to the cluster, it might be a
member of the recently proposed class of variable stars termed ``red
stragglers'' (Albrow et al.\ 2001) or ``sub-subgiant stars'' (Mathieu
et al.\ 2003). Thus far, the origin and evolutionary status of these
stars remains unknown.

In Fig.~\ref{lc:misc} we show the light curves of the cataclysmic
variables B7 and B8. In Fig.~\ref{lc:ecl} they are phased with cycle
lengths 17.6 and 11.3 days, respectively. Equally good fits to the B7
and B8 data are given by periods of 22.12 and 17.75 days, closer to
the cycle lengths of 25.41 and 17.73, reported by Mochejska, Stanek \&
Kaluzny (2003). More observations are required to firmly establish the
cycle lengths of these variables.

In Fig.~\ref{lc:rscvn} we show the light curves of three RS CVn type
binaries V9, V12 and V33, plotted separately for four time intervals:
2001 July 9 - August 1, 2002 May 14 - June 28, 2002 September 10 -
November 18 and 2003 May 5 - July 10. The shape of their light curves
varies with time, especially for V12 and V33.

\section{{\sc Conclusions}}
In this paper we have performed an extensive search for transiting
planets in the very old, populous, metal rich cluster NGC~6791. The
cluster was monitored for over 300 hours during 84 nights. We have not
detected any promising transiting planet candidates. Assuming a planet
frequency from radial velocity surveys, we estimate that we should
have detected 1.5 transiting planets with periods between 1 and 10
days, with our photometric precision and temporal coverage. The main
limitation on our detection efficiency was imposed by the photometric
precision. 

We have discovered 14 new variable stars in NGC 6791: nine eclipsing
binaries, four other periodic variables and one non-periodic variable,
bringing the total number of variables in this cluster to 111. We have
also presented high photometric precision light curves, spanning two
years, for all previously known variables. Many of them show changes
in light curve shape, i.e.\ V9, V12 and V33 (Fig.\ \ref{lc:rscvn}).
This phenomenon is most likely due to the evolution of magnetic spots
on the surface of these stars.

Transiting planets have proven to be more challenging to detect than
initially expected, as shown by the paucity of detections from the
many searches under way in open clusters (i.e.\ Bruntt et al.\ 2003;
UStAPS: Street et al.\ 2003; EXPLORE/OC: von Braun et al.\ 2004;
STEPSS: Burke et al.\ 2004) and in the Galactic field (i.e.~EXPLORE:
Mall{\' e}n-Ornelas et al.\ 2003; OGLE: Udalski et al.\ 2002a; STARE:
Alonso et al.\ 2003; HAT: Bakos et al.\ 2004\footnote{For a more
complete list of transiting planet searches, please refer to {\tt
http://star-www.st-and.ac.uk/\~{}kdh1/transits/table.html} and {\tt
http://www.obspm.fr/encycl/searches.html}}). To date, only six planets
have been discovered independently by transit searches, all of them in
the field, and five of those were initially identified by OGLE
(Udalski et al.\ 2002a, 2002b, 2002c, 2003; Alonso et al.\ 2004).

\acknowledgments{ We would like to thank the FLWO 1.2 m TAC for the
generous amount of time we were allocated to this project, the
anonymous referee for a prompt and useful report, Scott Gaudi and
Janusz Kaluzny for helpful discussions, Alceste Bonanos for her help
in obtaining some of the data, Andrzej Kruszewski for granting us
access to his light curve correction code and Peter McCullough for
advice on rejecting bad epochs.

This research has made use of the USNOFS Image and Catalogue Archive
operated by the United States Naval Observatory, Flagstaff Station
(http://www.nofs.navy.mil/data/fchpix/), the Digital Sky Survey,
produced at the Space Telescope Science Institute under
U.S. Government grant NAG W-2166, the SIMBAD database, operated at
CDS, Strasbourg, France and the WEBDA open cluster database maintained
by J.~C.~Mermilliod (http://obswww.unige.ch/webda/).

Support for BJM, GAB and JNW was provided by NASA through Hubble
Fellowship grants HST-HF-01155.02-A, HF-01170.01-A, HST-HF-01180.01-A
from the Space Telescope Science Institute, which is operated by the
Association of Universities for Research in Astronomy, Incorporated,
under NASA contract NAS5-26555. KZS acknowledges support from the
William F.~Milton Fund.}

\input{tab1.tex}

\input{tab2.tex}
\input{tab3.tex}

\input{tab4.tex}
\input{tab5.tex}
\input{tab6.tex}

\end{document}

%% file: tab1.tex
\begin{deluxetable}{lrrrrrrrrrrrrrr}
\tabletypesize{\footnotesize}
\tablewidth{0pt}
\tablecaption{Calibration Coefficients}
\tablehead{
\colhead{} & 
\multicolumn{4}{c}{V}& \colhead{}&
\multicolumn{4}{c}{V-R}& \colhead{}&
\multicolumn{4}{c}{R}\\
\colhead{chip} & 
\colhead{$a_1$}&\colhead{$a_2$}&\colhead{$a_3$}&\colhead{$rms$}&\colhead{}&
\colhead{$b_1$}&\colhead{$b_2$}&\colhead{$b_3$}&\colhead{$rms$}&\colhead{}&
\colhead{$c_1$}&\colhead{$c_2$}&\colhead{$c_3$}&\colhead{$rms$}}
\startdata
1 & 2.8989 & 0.0585 &0.1733 & 0.009 &
  & 0.2089 & 1.0941 &0.0305 & 0.016 &
  & 2.6916 &-0.0361 &0.1338 & 0.014\\
2 & 3.2527 & 0.0627 &0.1664 & 0.009 &
  & 0.3349 & 1.1002 &0.0178 & 0.010 &
  & 2.9206 &-0.0379 &0.1370 & 0.011\\
3 & 2.7596 & 0.0734 &0.1557 & 0.006 &
  & 0.1937 & 1.0952 &0.0430 & 0.006 &
  & 2.5674 &-0.0218 &0.1056 & 0.008\\
4 & 2.8950 & 0.0548 &0.1318 & 0.005 &
  & 0.2623 & 1.0823 &0.0011 & 0.015 &
  & 2.6339 &-0.0276 &0.1239 & 0.013\\
\enddata
\label{tab:cal}
\end{deluxetable}

%% file: tab2.tex
\begin{deluxetable}{lllcc}
\tabletypesize{\footnotesize}
\tablewidth{0pt}
\tablecaption{Parameter Range}
\tablehead{
\colhead{Parameter} & \colhead{min} & \colhead{max} & \colhead{step}
& \colhead{n$_{steps}$}}
\startdata
P (days)      &  1.05   &  9.85 &  0.200 &  45\\
$R_P$ ($R_J$) &  0.95   &  1.50 &  0.050 &  12\\
$T_0$         &  0.00   &  0.95 &  0.050 &  20\\
$\cos i$      &  0.0125 &  0.9875 &  0.025 &  40\\
\enddata
\label{tab:pars}
\end{deluxetable}

%% file: tab3.tex
\begin{deluxetable}{crrrrrrrr}
\tabletypesize{\footnotesize}
\tablewidth{0pt}
\tablecaption{Artificial transit test statistics}
\tablehead{
\colhead{test} & 
\multicolumn{2}{c}{all transits} &
\multicolumn{2}{c}{model} &
\multicolumn{2}{c}{marginal} &
\multicolumn{2}{c}{firm}\\
\colhead{type} & 
\colhead{N}&\colhead{\%} &
\colhead{N}&\colhead{\%} &
\colhead{N}&\colhead{\%} &
\colhead{N}&\colhead{\%}}
\startdata
 A & 43371 &   10.0 & 32406 &   74.7 &  8737 &   20.1 &  4323 &   10.0\\
 B & 43367 &   10.0 & 32380 &   74.7 &  9173 &   21.2 &  4659 &   10.7\\
 C & 43371 &   10.0 & 32406 &   74.7 &  5754 &   13.3 &  1998 &    4.6\\
\enddata
\label{tab:art}
\end{deluxetable}

%% file: tab4.tex
\begin{deluxetable}{rrrrrrrrlccccc}
\tabletypesize{\footnotesize}
\tablewidth{0pc}
\tablecaption{Eclipsing binaries in NGC 6791}
\tablehead{\colhead{ID} & \colhead{$\alpha_{2000}$ [h]} &
\colhead{$\delta_{2000}$ [$\circ$]} &\colhead{P [d]} & \colhead{$R_{max}$} &
\colhead{$V_{max}$} &\colhead{$A_R$} & \colhead{$A_V$}}
\startdata
  V22 & 19 20 18.7 & 37 30 29.8 &  0.2451 & 18.917 & 19.654  &  0.693 &  1.113 \\
  V01 & 19 20 47.6 & 37 44 32.0 &  0.2677 & 15.718 & 16.241  &  0.308 &  0.393 \\
  V23 & 19 20 19.0 & 37 47 16.0 &  0.2718 & 16.196 & 16.856  &  0.071 &  0.099 \\
  V02 & 19 21 17.5 & 37 46 00.2 &  0.2735 & 19.074 & 19.537  &  0.198 &  0.546 \\
  V24 & 19 19 58.5 & 37 35 44.0 &  0.2758 & 18.282 & 19.001  &  0.227 &  0.344 \\
  V25 & 19 19 42.3 & 37 42 48.1 &  0.2774 & 17.852 & 18.554  &  0.447 &  0.522 \\
  V06 & 19 21 02.7 & 37 48 48.9 &  0.2790 & 14.972 & 15.430  &  0.101 &  0.120 \\
  V26 & 19 20 44.9 & 37 33 42.6 &  0.2836 & 16.798 & 17.332  &  0.212 &  0.237 \\
  V05 & 19 20 46.5 & 37 48 47.8 &  0.3127 & 16.669 & 17.193  &  0.050 &  0.078 \\
  V03 & 19 21 15.8 & 37 46 09.7 &  0.3176 & 17.955 & 18.535  &  0.091 &  0.188 \\
  V04 & 19 20 54.2 & 37 48 23.8 &  0.3256 & 17.170 & 17.771  &  0.102 &  0.118 \\
  V27 & 19 20 10.7 & 37 38 56.5 &  0.3317 & 17.985 & 18.549  &  0.646 &  0.840 \\
 V101 & 19 21 05.6 & 37 38 25.3 &  0.3348 & 19.483 & 19.925  &  0.310 &  0.425 \\
 V102 & 19 19 31.0 & 37 32 16.0 &  0.3667 & 18.911 & 19.314  &  0.377 &  0.498 \\
  V28 & 19 19 43.8 & 37 35 30.2 &  0.3721 & 16.948 & 17.467  &  0.420 &  0.552 \\
  V40 & 19 19 39.0 & 37 37 01.0 &  0.3975 & 19.033 & 19.748  &  0.163 &  0.210 \\
  V29 & 19 21 17.3 & 37 45 05.2 &  0.4366 & 19.083 & 20.046  &  0.193 &  0.236 \\
  V41 & 19 20 51.0 & 37 48 24.7 &  0.4817 & 18.359 & 19.072  &  0.111 &  0.202 \\
 V103 & 19 20 35.6 & 37 35 45.0 &  0.6479 & 20.856 & \nodata &  0.779 & \nodata \\
  B04 & 19 21 12.9 & 37 45 51.3 &  0.7970 & 17.910 & 17.873  &  0.063 &  0.113 \\
  V11 & 19 20 33.3 & 37 48 16.6 &  0.8831 & 18.843 & 19.449  &  0.419 &  0.670 \\
 V104 & 19 20 43.3 & 37 34 40.6 &  1.0911 & 18.538 & 19.675  &  0.114 &  0.423 \\
  V33 & 19 20 39.8 & 37 43 54.4 &  1.1725 & 15.522 & 16.224  &  0.080 &  0.172 \\
  V30 & 19 19 43.0 & 37 30 06.9 &  1.1790 & 15.746 & 16.074  &  0.025 &  0.037 \\
  V80 & 19 21 06.5 & 37 47 27.8 &  1.2215 & 17.142 & 17.738  &  0.107 &  0.174 \\
 V105 & 19 20 39.1 & 37 33 36.2 &  1.2632 & 20.048 & 20.412  &  0.371 &  0.713 \\
 V106 & 19 21 10.7 & 37 45 31.6 &  1.4464 & 15.389 & 15.685  &  0.016 &  0.022 \\
  V43 & 19 20 39.6 & 37 38 30.7 &  1.5140 & 18.186 & \nodata &  0.075 & \nodata \\
  V12 & 19 20 42.9 & 37 50 56.5 &  1.5232 & 16.931 & 17.499  &  0.248 &  0.340 \\
 V107 & 19 21 18.2 & 37 45 41.8 &  1.6578 & 17.434 & 17.999  &  0.157 &  0.236 \\
  V32 & 19 20 27.6 & 37 47 14.2 &  2.0703 & 18.760 & 19.334  &  0.130 &  0.269 \\
  V34 & 19 20 09.2 & 37 44 10.7 &  2.4059 & 18.410 & 19.201  &  0.193 &  0.336 \\
  V36 & 19 19 56.4 & 37 34 12.6 &  2.6722 & 15.517 & 16.323  &  0.057 &  0.093 \\
 V108 & 19 21 09.4 & 37 49 24.5 &  2.8517 & 20.117 & \nodata &  0.870 & \nodata \\
  V09 & 19 20 47.9 & 37 46 37.4 &  3.1874 & 16.458 & 17.219  &  0.225 &  0.370 \\
  V37 & 19 21 18.2 & 37 51 07.0 &  3.2133 & 18.353 & 19.535  &  0.156 &  0.648 \\
  V35 & 19 20 44.1 & 37 30 42.8 &  3.2189 & 16.628 & 17.139  &  0.235 &  0.255 \\
  V31 & 19 21 02.5 & 37 47 09.3 &  3.3147 & 16.565 & 17.125  &  0.021 &  0.036 \\
 V109 & 19 20 33.8 & 37 47 37.4 &  3.7022 & 19.766 & \nodata &  0.760 & \nodata \\
  V38 & 19 21 03.7 & 37 46 05.9 &  3.8704 & 18.195 & 18.833  &  0.192 &  0.239 \\
  V60 & 19 21 00.7 & 37 45 45.0 &  7.4532 & 18.083 & 18.697  &  0.320 &  0.678 \\
  V20 & 19 20 54.3 & 37 45 34.7 &  7.4742 & 16.823 & 17.377  &  0.271 &  0.288 \\
  V39 & 19 21 00.5 & 37 38 22.8 &  7.6006 & 15.956 & 16.678  &  0.079 &  0.098 \\
  V14 & 19 20 51.7 & 37 45 24.8 & 10.9853 & 18.065 & \nodata &  0.073 & \nodata \\
  V18 & 19 20 49.4 & 37 46 09.2 & 17.6389 & 17.197 & \nodata &  0.433 & \nodata \\
  V61 & 19 19 42.9 & 37 29 07.4 & 19.3807 & 16.314 & 16.888  &  0.467 &  0.561 \\
\enddata
\label{tab:ecl}
\end{deluxetable}

%% file: tab5.tex
\begin{deluxetable}{rrrrrrrrlccc}
\tabletypesize{\footnotesize}
\tablewidth{0pc}
\tablecaption{Other periodic variables in NGC 6791}
\tablehead{\colhead{ID} & \colhead{$\alpha_{2000}$ [h]} &
\colhead{$\delta_{2000}$ [$\circ$]} &\colhead{P [d]} & \colhead{$\langle R\rangle$} &
\colhead{$\langle V\rangle$} &\colhead{$A_R$} & \colhead{$A_V$}}
\startdata
 V42 & 19 21 00.2 & 37 42 53.4 &  0.5064 & 18.954 & 19.597  &  0.035 &  0.044 \\
 V93 & 19 21 05.2 & 37 47 08.4 &  0.9941 & 16.473 & 16.925  &  0.003 &  0.004 \\
 V110 & 19 21 05.8 & 37 44 30.4 &  1.1329 & 17.342 & 17.828  &  0.005 &  0.006 \\
 V111 & 19 20 49.1 & 37 48 43.7 &  1.5487 & 19.672 & 20.555  &  0.033 &  0.030 \\
 V84 & 19 20 47.7 & 37 44 58.2 &  1.6258 & 18.989 & 19.836  &  0.021 &  0.027 \\
 V44 & 19 19 37.1 & 37 41 41.7 &  2.2544 & 17.782 & 18.410  &  0.013 &  0.019 \\
 V16 & 19 21 07.6 & 37 48 09.6 &  2.2664 & 17.276 & 17.850  &  0.019 &  0.024 \\
 V76 & 19 20 49.9 & 37 45 50.9 &  4.0924 & 17.585 & 18.270  &  0.030 &  0.034 \\
 V49 & 19 20 30.9 & 37 36 51.2 &  4.9923 & 15.266 & 15.802  &  0.007 &  0.007 \\
 V45 & 19 20 46.1 & 37 42 05.9 &  5.0883 & 16.463 & 17.083  &  0.007 &  0.010 \\
 V46 & 19 21 19.0 & 37 47 56.1 &  5.1287 & 17.930 & 18.688  &  0.033 &  0.037 \\
 V47 & 19 19 39.1 & 37 32 10.8 &  5.6066 & 18.737 & 19.978  &  0.024 &  0.011 \\
 V91 & 19 21 00.5 & 37 48 40.6 &  5.6411 & 17.611 & 18.097  &  0.004 &  0.004 \\
 V48 & 19 21 07.5 & 37 43 06.6 &  5.8019 & 17.003 & 17.558  &  0.018 &  0.013 \\
 V50 & 19 20 35.2 & 37 31 04.3 &  5.8812 & 16.055 & 16.544  &  0.011 &  0.008 \\
 V89 & 19 20 56.6 & 37 46 36.2 &  6.1577 & 18.226 & 19.078  &  0.025 &  0.053 \\
 V17 & 19 20 38.9 & 37 49 04.5 &  6.3656 & 17.279 & 17.949  &  0.020 &  0.020 \\
 V52 & 19 21 20.9 & 37 46 19.2 &  6.9933 & 17.016 & \nodata &  0.005 & \nodata \\
 V51 & 19 21 12.2 & 37 44 54.7 &  7.0315 & 19.257 & 19.971  &  0.032 &  0.025 \\
 V77 & 19 20 52.9 & 37 46 36.9 &  7.1810 & 16.206 & 16.744  &  0.003 &  0.004 \\
 V53 & 19 21 00.8 & 37 44 35.4 &  7.1822 & 18.257 & 18.803  &  0.011 &  0.006 \\
 V83 & 19 20 46.4 & 37 44 14.1 &  7.2915 & 18.683 & 19.392  &  0.013 &  0.017 \\
 V82 & 19 20 39.7 & 37 47 36.0 &  7.4983 & 18.522 & 19.064  &  0.018 &  0.008 \\
 V54 & 19 21 18.7 & 37 43 36.4 &  8.3141 & 15.929 & 16.524  &  0.010 &  0.010 \\
 V98 & 19 20 56.5 & 37 45 38.7 &  8.3405 & 16.418 & \nodata &  0.003 & \nodata \\
 V112 & 19 20 04.2 & 37 48 33.4 &  8.7902 & 14.947 & 15.467  &  0.003 &  0.002 \\
 V95 & 19 20 43.1 & 37 47 32.5 &  9.6832 & 18.509 & 19.147  &  0.014 &  0.010 \\
 V65 & 19 20 52.5 & 37 47 30.5 & 11.1091 & 15.645 & 16.272  &  0.004 &  0.006 \\
 V56 & 19 20 45.3 & 37 45 48.8 & 12.3832 & 16.518 & 17.081  &  0.006 &  0.003 \\
 V57 & 19 20 57.9 & 37 31 07.0 & 13.1536 & 17.460 & 18.359  &  0.010 &  0.007 \\
 V58 & 19 21 14.5 & 37 48 04.4 & 13.2597 & 17.028 & 17.544  &  0.015 &  0.016 \\
 V113 & 19 20 34.9 & 37 48 14.8 & 13.5908 & 16.995 & 17.568  &  0.009 &  0.013 \\
 V97 & 19 20 49.2 & 37 49 14.8 & 13.6206 & 15.813 & 16.517  &  0.003 &  0.004 \\
 V59 & 19 20 21.5 & 37 48 21.9 & 13.8331 & 17.221 & 17.781  &  0.056 &  0.049 \\
 V64 & 19 21 11.4 & 37 29 55.4 & 14.2427 & 15.882 & 16.473  &  0.005 &  0.012 \\
 V81 & 19 20 49.7 & 37 48 08.7 & 16.6182 & 16.373 & 16.905  &  0.003 &  0.005 \\
 V100 & 19 21 01.8 & 37 45 41.9 & 23.9468 & 16.544 & 17.151  &  0.024 &  0.021 \\
 V66 & 19 21 08.4 & 37 44 55.2 & 50.4976 & 15.372 & 16.119  &  0.068 &  0.074 \\
 V71 & 19 21 10.5 & 37 43 24.8 & 51.9430 & 16.563 & 17.291  &  0.057 &  0.067 \\
 V67 & 19 21 03.7 & 37 48 03.7 & 66.7944 & 16.180 & 16.959  &  0.056 &  0.071 \\
\enddata
\label{tab:pul}
\end{deluxetable}

%% file: tab6.tex
\begin{deluxetable}{rrrrrrrlccc}
\tabletypesize{\footnotesize}
\tablewidth{0pc}
\tablecaption{Miscellaneous variables in NGC 6791}
\tablehead{\colhead{ID} & \colhead{$\alpha_{2000}$ [h]} &
\colhead{$\delta_{2000}$ [$\circ$]} & \colhead{$R_{max}$} &
\colhead{$V_{max}$} &\colhead{$A_R$} &\colhead{$A_V$}}
\startdata
 B07 & 19 21 07.4 & 37 47 56.5 & 17.448 & 17.581  &  0.819 &  0.919  \\
 B08 & 19 20 35.7 & 37 44 52.3 & 18.820 & 18.716  &  1.896 &  3.403  \\
 V10 & 19 21 11.8 & 37 47 58.1 & 18.922 & 19.642  &  0.103 &  0.435  \\
 V21 & 19 20 57.3 & 37 45 36.9 & 16.997 & 17.547  &  0.016 &  0.041  \\
 V62 & 19 21 03.0 & 37 43 51.8 & 18.618 & 19.189  &  0.105 &  0.214  \\
 V63 & 19 19 40.0 & 37 29 45.1 & 16.317 & 17.058  &  0.024 &  0.066  \\
 V70 & 19 20 32.2 & 37 44 21.0 & 99.999 & 14.722  &  0.000 &  0.427  \\
 V74 & 19 21 07.2 & 37 44 34.9 & 99.999 & 14.660  &  0.000 &  0.025  \\
 V75 & 19 20 47.9 & 37 45 58.8 & 16.829 & 17.374  &  0.021 &  0.053  \\
 V79 & 19 20 55.2 & 37 46 39.7 & 18.034 & 18.631  &  0.083 &  0.261  \\
 V86 & 19 20 50.1 & 37 48 31.7 & 18.846 & 19.478  &  0.134 &  0.550  \\
 V87 & 19 20 52.8 & 37 44 58.8 & 17.637 & 18.189  &  0.038 &  0.094  \\
 V90 & 19 20 58.9 & 37 44 47.1 & 17.608 & 18.159  &  0.039 &  0.148  \\
 V94 & 19 20 42.5 & 37 44 36.9 & 17.023 & 17.563  &  0.034 &  0.081  \\
 V99 & 19 20 57.1 & 37 48 12.1 & 16.866 & 17.527  &  0.025 &  0.045  \\
 V114 & 19 20 00.0 & 37 48 44.7 & 16.882 & 17.601  &  0.034 &  0.057  \\
\enddata
\label{tab:misc}
\end{deluxetable}